\documentclass[twocolumn,showpacs,preprintnumbers,amsmath,amssymb]{revtex4}

\usepackage{graphicx,textcomp}
\usepackage{dcolumn}
\usepackage{bm}

\begin{document}


\title{High-resolution spectroscopy of triplet states of Rb$_2$ by femtosecond pump-probe photoionization of doped helium nanodroplets}

\author{M. Mudrich$^1$}
\author{Ph. Heister$^1$}
\author{T. Hippler$^1$}
\author{Ch. Giese$^1$}
\author{O. Dulieu$^2$}
\author{F. Stienkemeier$^1$}

\affiliation{$^1$Physikalisches Institut, Universit\"at Freiburg, 79104 Freiburg, Germany}
\affiliation{$^2$Laboratoire Aim\'e Cotton, CNRS, Universit\'e Paris-Sud, B\^atiment 505, 91405 Orsay, France}

\date{\today}

\begin{abstract}
\noindent
The dynamics of vibrational wave packets in triplet states of rubidium dimers (Rb$_2$) formed on helium nanodroplets are studied using femtosecond pump-probe photoionization spectroscopy. Due to fast desorption of the excited Rb$_2$ molecules off the droplets and due to their low internal temperature, wave packet oscillations can be followed up to very long pump-probe delay times $\gtrsim 1.5\,$ns. In the first excited triplet state $(1)^3\Sigma_g^+$, full and fractional revivals are observed with high contrast. Fourier analysis provides high-resolution vibrational spectra which are in excellent agreement with \textit{ab initio} calculations.

\end{abstract}

\maketitle
\section{Introduction}
\label{sec:intro}
The dynamics of vibrational wave packets (WP) has been investigated for a large variety of molecular systems ranging from simple diatomics to biomolecules~\cite{Zewail:1994,Manz:1995,Chergui:1995}. Alkali diatomic molecules were among the first molecules to be studied due to their strong electronic transitions in the visible range of the spectrum. Low ionization potentials make alkali dimers accessible to photoionization (PI) using comparatively low laser intensities. Besides, potential energy curves can be calculated with high precision, thus facilitating the interpretation of spectroscopic data (see \textit{e.\,g.}~\cite{Aymar:2005} and references therein).

A number of interesting phenomena have been studied on these simple molecules, \textit{e.\,g.} WP propagation in spin-orbit-coupled states~\cite{Rutz:1996,Zhang:2003}, the competition of different ionization pathways~\cite{Riedle:1996,Nicole:1999}, fractional revivals of vibrational WPs~\cite{Rutz:1997,Vrakking:1996}, and isotope-selective ionization~\cite{LindingerPRL:2004}. Detailed insights into the vibrational dynamics have been obtained by applying new experimental techniques such as photo-electron spectroscopy~\cite{Wollenhaupt:2005}, and optimal control schemes using shaped laser pulses~\cite{Bung:2004,Baumert:2005}.

Besides, valuable spectroscopic information can be gained from time-resolved experiments~\cite{Rutz:1996,Oldenbourg:2000,Zhang:2003}. The extraction of high resolution spectra and potential energy curves from ultrafast laser experiments has been discussed in detail by Gruebele and Zewail, using molecular iodine I$_2$ as a model system~\cite{Gruebele:1992}.

Alkali dimers have attracted new interest in the past years due to the recent advances in the formation of ultracold molecules out of ultracold atomic ensembles by means of Feshbach resonances~\cite{Chin:2009} and photo-association~\cite{Masnou:2001,Jones:2006}. These studies require the knowledge of molecular spectra with great precision~\cite{Lozeille:2006,Danzl:2009}.
However, using conventional spectroscopy of molecules in the gas-phase which are formed in the singlet ground state, triplet transitions are more difficult to interrogate.

Pump-probe spectroscopy of alkali dimers attached to He nanodroplets has been reported before by our group~\cite{Claas:2006,Claas:2007}. Helium (He) nanodroplets are widely used as a nearly ideal cryogenic matrix for spectroscopy of embedded atoms, molecules, and clusters~\cite{Toennies:2004,Stienkemeier:2006}. Alkali atoms and molecules represent a
particular class of dopant particles characterized by their extremely weak binding to He
droplets with binding energies of about 10\,K~\cite{Mayol:2005}. From both theory and experiment it
is known that alkali dimers reside in bubble-like structures on the surface of He
droplets~\cite{Dalfovo:1994,Ancilotto:1995,Stienkemeier2:1995,Mayol:2005}.
Upon laser excitation, alkali atoms and molecules mostly desorb off the droplet surface as
a consequence of repulsive interactions with the He surroundings induced by the change of electronic and vibrational states~\cite{Stienkemeier:1996,Callegari:1998,Buenermann:2007}.

The influence of the He droplet on the vibrational dynamics of attached molecules has been studied using potassium dimers (K$_2$) in singlet states~\cite{Claas:2006}. Transient modulations of both amplitudes and frequencies of vibrational frequency components were observed, from which the time constant for the desorption dynamics was estimated to range between 3 and 8\,ps. WP propagation in triplet states was observed for the first time using sodium dimers (Na$_2$) attached to He nanodroplets~\cite{Claas:2007}. However, the triplet-state dynamics did not reveal any influence of the He matrix on the WP dynamics. This may be due to considerably shorter desorption times of alkali dimers in triplet states as a result of more efficient coupling to the surrounding He as compared to singlet states. Bovino~\textit{et al.} have calculated enhanced vibrational quenching cross sections by orders of magnitude for vibrational states of the lowest triplet state $a^3\Sigma_u^+$ of Li$_2$ as compared to the singlet ground state~\cite{Bovino:2008}. Furthermore, recent studies of the cw excitation spectra of Rb$_2$ on He droplets provides indications that triplet dimers are oriented parallel to the He surface and thus may couple more efficiently to surface excitations of the He droplet~\cite{Auboeck:2007}.

This work presents a systematic study of the pump-probe dynamics of Rb$_2$ formed on He nanodroplets in triplet states in the entire tuning range of the Titanium:Sapphire (Ti:Sa) laser between $\lambda=700$\,nm and $\lambda=1030\,$nm. Only weak perturbations of the vibrational dynamics by the He matrix are observed at short delay times $t\lesssim 10\,$ps associated with the desorption dynamics. Thus, the dynamics probed in our experiment corresponds mostly to the one of free Rb$_2$ molecules in the gas-pase. Due to the low internal temperature of Rb$_2$ formed on He droplets, exceptionally long coherence times of the vibrational WP oscillations are observed. This allows to obtain high-resolution Fourier spectra of the vibrational frequencies of triplet states of Rb$_2$, which may be of considerable interest in the context of the formation of ultracold Rb$_2$ molecules~\cite{Lozeille:2006,Lang:2008}.

 The paper is organized as follows: Sec.~\ref{sec:WP} gives a short introduction into the theory of vibrational WPs in anharmonic potentials. Sec.~\ref{sec:experimental} outlines the experimental arrangement. The discussion of experimental results begins with the presentation of PI and beam depletion spectra in Sec.~\ref{sec:PIspectra}. It is followed by a more detailed discussion of the temporal and spectral data measured in different laser wavelength ranges corresponding to three different electronic transitions (Sec.~\ref{sec:leftpeak}-\ref{sec:rightpeak}). A quantitative analysis of the pump-probe spectra measured around $\lambda=1000\,$nm including the comparison with theoretical potential curves is given in Sec.~\ref{sec:analysis}.

\begin{figure}[t!]
\begin{center}
{
\includegraphics[width=0.51\textwidth]{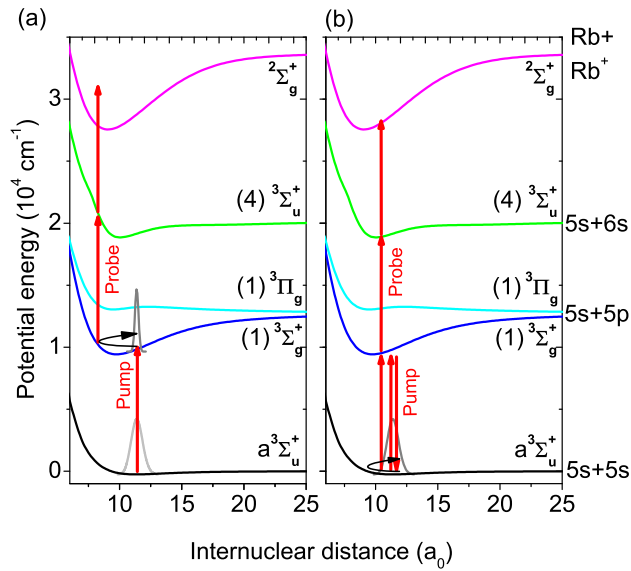}}
\caption{Selected triplet potential energy curves of neutral Rb$_2$ and of the Rb$_2^+$ ionic ground state relevant to the present study. The arrows in (a) and (b) indicate two different pathways leading to ionization: (a) Creation of a wave packet in the first excited state followed by resonant two-photon ionization; (b) Creation of a wave packet in the lowest triplet state by impulsive Raman scattering followed by three-photon ionization.
}
\label{fig:Pots}
\end{center}
\end{figure}
\section{Vibrational wave packet dynamics}
\label{sec:WP}
The dynamics of vibrational WPs is studied using the pump-probe technique. The principle of operation is the following. The first pump pulse creates a vibrational WP, $\psi(x,t)=\sum_n a_n|n\rangle \exp -iE_nt/\hbar$, \textit{i.\,e.} a coherent superposition of vibrational states of a bound electronic state, as illustrated in Fig.~\ref{fig:Pots} for the case of Rb$_2$. Here, $|n\rangle$ denotes the set of vibrational eigenstates with energies $E_n$, and $a_n$ are constant coefficients. A second fs laser pulse probes the WP through excitation to a higher electronic state $|f\rangle$ which is detected with time-independent methods, \textit{i.\,e.} by measuring the yield of photo-ions. In the scheme shown in Fig.~\ref{fig:Pots} (a) the probe step consists of a resonant two-photon transition towards the ionic ground state $^2\Sigma_g^+$ via the $4^3\Sigma_u^+$-state of Rb$_2$. The ion signal $S(t)$ is detected as a function of the time delay $t$ between the two pulses.

According to the Franck-Condon principle, the WP is detected near one or a few internuclear separations (Condon points), at which the energy difference between upper and lower potential curves match the laser energy. As the WP swings back and forth inside the potential well, excitation to the ionic state becomes maximum each time the WP passes a Condon point. This causes the detector signal to be modulated periodically. In case there is only one Condon point which coincides with one classical turning point, the transition probability to the ionic state is particularly high due to the favorable overlap between wave functions in lower and upper states (Franck-Condon factor, FCF). This condition is fulfilled in the scheme shown in Fig.~\ref{fig:Pots} (a). In this case, the period of the ion signal approximately equals the classical frequency of the molecular vibration, $\omega_c$. The measured signal $S(t)=|P(t)|^2$ is given by $P(t)=\langle f|\vec\mu\cdot\vec E|\psi(x,t)\rangle=\sum_nb_n\exp-iE_n t/\hbar$, where $\langle f |$ represents the final ionic state, $\vec\mu$ the transition dipole moment, $\vec E$ the electric field vector of the probe laser, and $b_n=a_n\langle f|\vec\mu\cdot\vec E|n\rangle$.  Therefore the signal $S(t)=\sum_{n,m}b_nb_m\cos\left[(E_n-E_m)t/\hbar\right]$ is composed of beat frequencies between all pairs of energy levels that make up the WP. The lowest frequencies are given by the beats between neighboring vibrational levels $|n\rangle$ and $|m\rangle$, where $n=m\pm 1$. These frequencies approximately match $\omega_c$. Beat notes between further-separated levels, $n=m\pm\ell$, where $\ell >1$, generate higher harmonics of $\omega_c$~\cite{Vrakking:1996}.

WP propagation in Morse-like anharmonic potentials has been studied theoretically by Vetchinkin \textit{et al.}~\cite{Vetchinkin:1993,Vetchinkin:1994,Vrakking:1996}. The energy levels of a Morse potential $V(R)=E_D\,\left(1-\exp\left[-a(R-R_e)\right]\right)^2$ are given by $E_v=\hbar\omega \left[(v+1/2)-x_e (v+1/2)^2\right]$. Here, $\omega=4E_Dx_e/\hbar=a\sqrt{2 E_D/\mu}$ stands for the harmonic oscillator frequency where $E_D$ is the dissociation energy, $x$ denotes the anharmonicity constant, $a$ is the range-parameter of the Morse potential, and $\mu=m_{\mathrm{Rb}}/2$ is the reduced mass of the Rb$_2$ molecule.

In anharmonic potentials the classical-like behavior decays with time due to dispersion of the initially created WP. In the Morse potential, however, the classical behavior partly and fully recurs at certain times, known as full and fractional revivals, respectively. The time at which the WP is fully reconstituted with its original phase is given by $T_{rev}=2h/(\Delta E_n-\Delta E_m)=2\pi/(\omega x_e)$ for any $m=n+1$, where $\Delta E_n=E_n-E_{n-1}$. Fractional revivals occur at times $t/T_{rev}=p/q$, where $p/q$ is an irreducible fraction of integers~\cite{Averbukh:1991}. It can be shown that the full revivals of the WP that occur at $t=T_{rev}/2,\,T_{rev},\,3T_{rev}/2,\dots$ result from lowest-frequency beats ($\ell =1$), while half-period revivals at $t=T_{rev}/4,\,T_{rev}/2,\,3T_{rev}/4,\dots$ result from second harmonic beats ($\ell =2$), etc.~\cite{Vrakking:1996}.

\section{Experimental}
\label{sec:experimental}
\begin{figure}[t!]
\begin{center}
{
\includegraphics[width=0.48\textwidth]{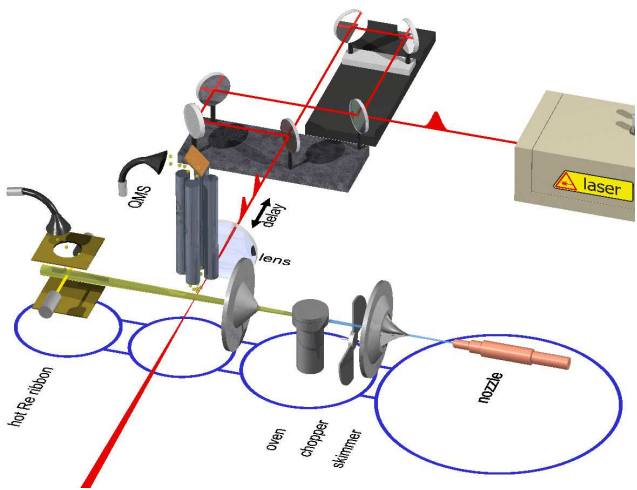}}
\caption{Schematic overview of the experimental setup. The upper part shows the optical system to produce femtosecond pulse pairs, the lower part shows the He droplet beam line.
}
\label{fig:experiment}
\end{center}
\end{figure}
The experimental setup is composed of two main parts, as schematically represented in Fig.~\ref{fig:experiment}. A molecular beam line produces He droplets doped with Rb$_2$ and an optical system supplies pairs of fs pulses with variable time delay. The beam line consists of four differentially pumped vacuum chambers. In the source chamber a continuous beam of He droplets is formed in a supersonic expansion of high purity $^4$He gas out of a cold nozzle ($T\approx 17$\,K, diameter $d=5\,\mu$m) at high pressure ($p\approx 50\,$bar). Under these conditions He droplets contain around 8000 He atoms~\cite{Toennies:2004}. The source chamber is pumped by three turbo pumps (1000\,l/s each) backed by a roots and rotary vane pump.

The beam enters the oven chamber through a commercial skimmer with a diameter of 0.4\,mm to pick up two Rb atoms out of atomic Rb vapor inside a pickup cell. The number of collisions of the droplets with free dopant atoms
inside the vapor cell is controlled by the oven temperature. Given the flight distance of
the droplets of 1\,cm inside the vapor cell the Rb reservoir is
heated to a temperature $T\approx $380\,K to achieve highest probability for pick-up of
two dopant atoms per droplet. Due to their high mobility on the superfluid He droplets the
picked-up atoms bind together to form dimers which are weakly bound in dimples on the
droplet surface. Collisional as well as binding energy is
dissipated by the He droplet through evaporation of He atoms, which may also cause the
desorption of the alkali dimers from the droplets. Since the amount of internal energy
released upon formation of ground state ($X^1\Sigma_g^+$) dimers greatly exceeds the one
released upon formation of dimers in the lowest triplet state $a^3\Sigma_u^+$, the later
has a higher chance to remain attached to the droplet. This leads to an enrichment of the
droplet beam with high-spin dimers and clusters compared to covalently bound systems
\cite{Higgins:1998,Stienkemeier:2001}. The alkali-He droplet complex eventually equilibrates at the terminal temperature of pure
He droplets of 380\,mK. Thus, only the lowest vibrational state, $v=0$, and a few
rotational states ($J\lesssim 10$) are populated which provides well-defined starting conditions for
pump-probe experiments.

In the third vacuum chamber the doped droplet beam is intersected by the
laser beam at right angle inside the detection volume of a commercial quadrupole mass spectrometer. The fs laser pulses are generated by a commercial mode-locked Ti:Sa laser (Chameleon, Coherent)
at 80\,MHz repetition rate and at an average
power output between 1.5 and 3.2\,W depending on the wavelength. The pulses have a duration of $\approx$160 fs and a full spectral bandwidth at half maximum (FWHM) of $\Delta\omega_{las}^{FWHM}\approx 80\,$cm$^{-1}$. The beam is sent through a Mach-Zehnder interferometer to split one pulse into two time-delayed pulses with equal intensity and phase. The time delay between the pulses is controlled by a commercial translation stage. In the experiments time increments of 30\,fs are used. Behind the interferometer the laser beams are recombined with parallel polarizations and focussed into the doped He droplet beam using a lens of 150\,mm focal length. The $1/e^2$-beam diameter in the focus is 41\,$\mu$m which leads to a pulse peak intensity between 0.5 and 4\,GW/cm$^2$ depending on the wavelength of the laser. The fourth chamber contains a Langmuir-Taylor surface ionization detector that is used to check the doping level and to perform beam depletion measurements~\cite{Callegari:1998,Stienkemeier:2000}.

\section{photoionization spectra}
\label{sec:PIspectra}
\begin{figure}[t!]
\begin{center}
{
\includegraphics[width=0.51\textwidth]{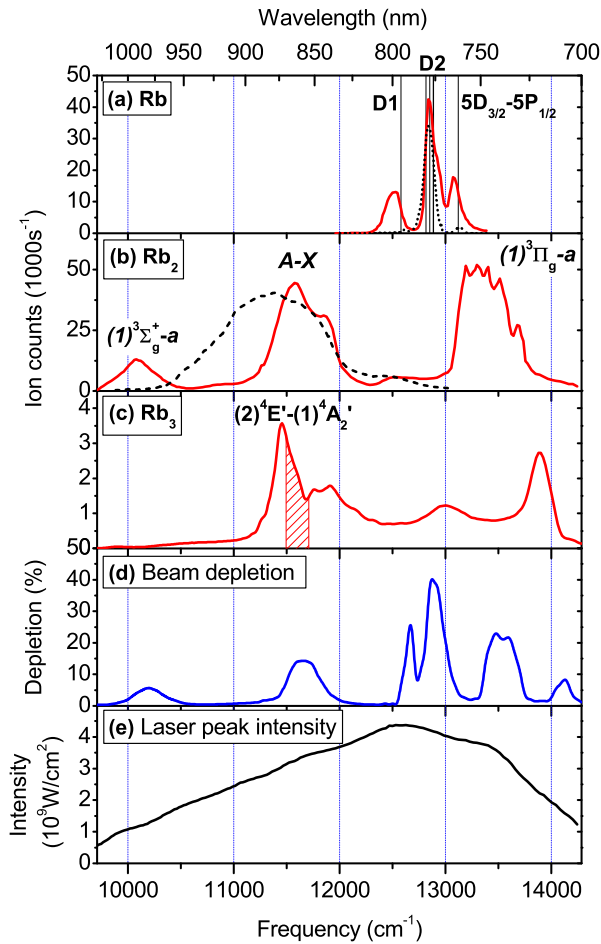}}
\caption{Comparison of fs photo-ionization spectra of Rb monomers, dimers and trimers attached to helium nanodroplets ((a)-(c)) with beam-depletion spectroscopy (d). The dashed lines in (a) and (b) are spectra of Rb and Rb$_2$, respectively, in the gas-phase. The shaded area in (c) indicates the absorption band $(2)^4E'\leftarrow (1)^4A_2'$ of Rb$_3$ measured by cw-spectroscopy~\cite{Nagl_prl:2008,Nagl_jcp:2008}. The spectra are not corrected for varying laser intensity (e).
}
\label{fig:PIspectra}
\end{center}
\end{figure}
In a first series of measurements, PI spectra are recorded with mass selection of the photo-ions using only one fs laser pulse at varying laser wavelength. PI spectra of Rb atoms, Rb$_2$ dimers and Rb$_3$ trimers attached to He nanodroplets are shown as solid lines in Fig.~\ref{fig:PIspectra} (a)-(c). For comparison, we have recorded PI spectra of Rb and Rb$_2$ in the gas-phase (dashed lines in Fig.~\ref{fig:PIspectra} (a) and (b)). Note that all spectra shown in Fig.~\ref{fig:PIspectra} are raw data that do not account for varying laser intensity (Fig.~\ref{fig:PIspectra} (d)). The gas-phase spectrum of Rb is dominated by a broad maximum around the D2-line (5P$_{3/2}\leftarrow$5S$_{1/2}$) at 780\,nm. The fs PI signal at that wavelength is additionally enhanced by the transition 5D$_{3/2\,[5/2]}\leftarrow$5P$_{3/2}$ which is blue-detuned from the D2-line by only 67.4\,cm$^{-1}$ [70.4\,cm$^{-1}$] and therefore simultaneously near-resonant given the bandwidth of the fs laser. A small peak around the transition 5D$_{3/2}\leftarrow$5P$_{1/2}$ is also visible, reflecting resonance-enhanced three-photon ionization (RE3PI) via the 5P$_{1/2}$ and 5D$_{3/2}$ states. Interestingly, the D1-transition (5P$_{1/2}\leftarrow$5S$_{1/2}$) is not observed as increased PI signal, presumably due to missing further resonance conditions for the transition up into the ionic state.

For PI of Rb attached to the surface of He nanodroplets, however, the situation is different. Here, a broad feature red-shifted to the D1-line is measured and the peak close to the 5D$_{3/2}\leftarrow$5P$_{1/2}$-transition is also enhanced. The difference between gas-phase and He droplet spectra is attributed to the interaction between the He environment and the Rb dopant which leads to shifting and broadening of atomic transitions~\cite{Buenermann:2007}. These matrix-induced perturbations are expected to be particularly pronounced for highly excited electronic states of Rb due to their extended orbitals and higher polarizabilities. Thus, apparently resonant-multi-photon ionization conditions are met at the wavelength of the D1-line of Rb attached to He droplets.

The PI spectra recorded at the mass of Rb$_2$ are depicted in Fig.~\ref{fig:PIspectra} (b). In order to assign the observed features of the spectrum of Rb$_2$ on He droplets (red line) we have also recorded the PI spectrum of gas-phase Rb$_2$ dimers formed inside the doping-cell heated to 380\,K by three-body collisions (dashed line). Gas-phase Rb$_2$ molecules are expected to occupy ro-vibrational states of the electronic ground state X$^1\Sigma_g^+$ according to their internal temperature. In the entire tuning range of the laser we observe one broad feature which is assigned to the $A^1\Sigma_u^+\leftarrow X^1\Sigma_g^+$-system and is analyzed in Sec.~\ref{sec:singlet}. In contrast, the He droplet spectrum reveals 3 broad peaks. This qualitative discrepancy between gas-phase and He droplet spectra reflects the different formation mechanisms of Rb$_2$ molecules out of single Rb atoms. In contrast to strongly bound Rb$_2$ formed in the hot vapor of the doping cell, the association of alkali dimers on the surface of He droplets favors weakly-bound states, \textit{i.\,e.} the lowest triplet state $a^3\Sigma_u^+$. We therefore assign the additional features showing up in the droplet spectrum around 1000\,nm and around 750\,nm to the triplet systems $(1)^3\Sigma_g^+\leftarrow a^3\Sigma_u^+$ and $(1)^3\Pi_g\leftarrow a^3\Sigma_u^+$, as detailed in sections~\ref{sec:leftpeak} and \ref{sec:rightpeak}.

In order to assign the center feature of the Rb$_2$ spectrum around 850\,nm we have also measured the PI spectrum of Rb$_3$ trimers formed on He droplets, shown in Fig.~\ref{fig:PIspectra} c). Fragmentation of Rb$_3^+ \rightarrow$Rb$_2^+$+Rb upon PI may cause a crosstalk of the Rb$_3^+$ ion signal onto the Rb$_2^+$ signal and thus perturb the Rb$_2$ PI spectrum. Rb$_3^+$ photo-ions are detected nearly in the entire scan range of the laser.
High-spin Rb$_3$ trimers formed on He droplets have been studied using cw spectroscopy~\cite{Nagl_prl:2008,Nagl_jcp:2008}. In the spectral range of interest, one electronic transition was identified ($(2)^4E'\leftarrow (1)^4A_2'$) in the range between 11500 and 11700\,cm$^{-1}$, indicated by the shaded area in Fig.~\ref{fig:PIspectra} (c). Indeed, the He droplet PI spectra measured on Rb$_2$ and Rb$_3$ masses in the range between 820 and 900\,nm show qualitative resemblance. This suggests that fragmentation of Rb$_3$ contributes to the Rb$_2$ spectrum. Besides, in this spectral range the Rb$_2$-He droplet spectrum partly coincides with the system $A\leftarrow X$ measured in the gas-phase. Therefore we conclude that presumably both the singlet transition $A\leftarrow X$ of Rb$_2$ as well as fragmentation of Rb$_3$ contribute to the PI spectrum measured at the mass of Rb$_2$.

Finally, Fig.~\ref{fig:PIspectra} (d) shows the beam depletion (BD) spectrum recorded using Langmuir-Taylor surface ionization detection. This measurement is performed using the unfocused fs laser beam to maximize the spatial overlap of the laser with the droplet beam. Thus, laser intensity is lower by about a factor 10$^{-3}$ as compared to the PI measurements mentioned above. We take as depletion signal the relative drop of surface ionization count rate when shining in the fs laser beam in proportion to the signal measured with blocked laser beam. Nearly all of the spectral features of the PI spectra recorded on masses of Rb, Rb$_2$ and Rb$_3$ are qualitatively reproduced. This clearly shows that the PI spectra stem from rubidium oligomers attached to He nanodroplets. The fact that the exact shapes and relative intensities of individual peaks differ from the ones seen in the PI spectra is mainly due to the much lower laser intensity. As a consequence, only the lowest lying transitions are expected to contribute to the signal, whereas the PI signal involves multi-photon resonances. Even though the detailed comparison of PI and BD spectra may provide new insight into He matrix-induced shifts and broadenings of higher lying electronic levels, it lies beyond the scope of this paper. In the following, the vibrational WP dynamics in 3 electronic states of Rb$_2$ are investigated using the one-color pump-probe technique. Wave packet oscillations of Rb$_3$ are also observed at laser wavelength 850\,nm and will be presented in a forthcoming publication.

\section{Excitation around 1000\,\lowercase{nm}}
\label{sec:leftpeak}
\begin{figure}[t!]
\begin{center}
{
\includegraphics[width=0.51\textwidth]{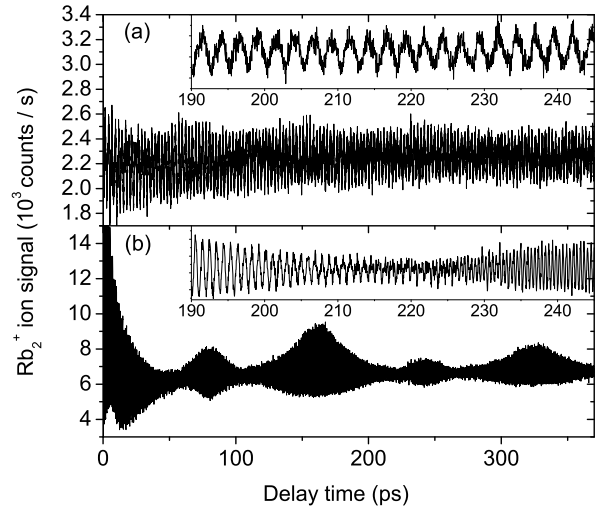}}
\caption{Experimental yield of Rb$^+_2$ photo-ions as a function of delay time between pump and probe laser pulses recorded at laser wavelength 970\,nm (a) and 1006\,nm (b). The insets show detailed views in the time range of the transition from full to half-period fractional revivals of the wave packet dynamics in the $(1)^3\Sigma_g^+$-state.
}
\label{fig:ppscan}
\end{center}
\end{figure}
The most pronounced WP oscillations are observed at a laser wavelength around $\lambda =$1000\,nm. Fig.~\ref{fig:ppscan} (a) displays the pump-probe transient measured at $\lambda=970$\,nm. The inset shows a detailed view of the same data in the delay time range between $t=190$ and $250$\,ps. The data nicely resolve WP oscillations with a period $T_a\approx 2.5$\,ps corresponding to the WP dynamics in the lowest triplet state $a^3\Sigma_u^+$. The oscillation amplitude $A_a$ slowly decreases but shows no periodic modulation. In contrast, the WP oscillations recorded at $\lambda=1006$\,nm, corresponding to dynamics in the first excited state $(1)^3\Sigma_g^+$, are strongly amplitude-modulated (Fig.~\ref{fig:ppscan} (b)). The period is $T_{\Sigma g}\approx 0.95$\,ps. After an initial decrease of the oscillation amplitude $A_{\Sigma g}$ in the time range between 0 and 50\,ps the signal builds up again to form a local maximum at about 80\,ps which corresponds to the first half-period fractional revival. The first full revival occurs at about 160\,ps. Up to about 10 full revivals are observed at delay times up to $\gtrsim 1.5\,$ns. The data shown in the inset illustrate the transition from the full revival to one-third and one-half period fractional revivals of the WP dynamics in the $(1)^3\Sigma_g^+$-state. The initial phase $\phi_{0,\,1006\,\mathrm{nm}}$ of the WP oscillation at $\lambda =1006\,$nm can be obtained from the extrapolation to $t=0$ of the linear fit of the positions of signal maxima as a function of delay time between $t=0$ and $t=20\,$ps. The resulting value $\phi_{0,\,1006\,\mathrm{nm}}=0.45\cdot 2\pi$ is consistent with the excitation scheme depicted in Fig.~\ref{fig:Pots} (a), which illustrates the propagation of a WP created at the right turning point of the $(1)^3\Sigma_g^+$-state potential and probed at the left turning point after 1/2 oscillation period. In addition to the periodic modulation due to revivals, the oscillation amplitude $A_{\Sigma g}$ is damped following roughly an exponential decay behavior with decay time constants varying between 100 and 700\,ps depending on the excitation laser wavelength. This slow damping is attributed to contributions of different rotational states to the measured transients, which have slightly shifted vibrational frequencies due to rotation-vibration coupling~\cite{Gruebele:1992}. Since the rotational temperature of the molecules formed on He nanodroplets is low compared to gas-cell or standard molecular beam experiments, though, this vibrational dephasing is very slow.


\begin{figure}[t!]
\begin{center}
{
\includegraphics[width=0.51\textwidth]{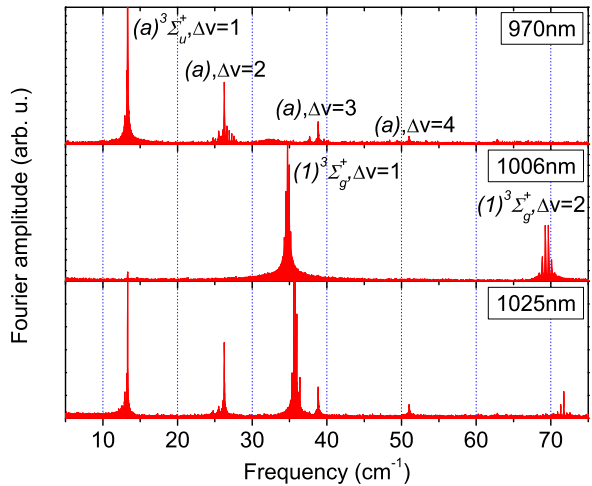}}
\caption{Fourier transforms of wave packet oscillations at different laser wavelengths. At laser wave length $\lambda=970$\,nm the spectrum is dominated by vibrations in the lowest triplet state $a^3\Sigma_u^+$, at $\lambda=1006$\,nm only the excited $(1)^3\Sigma_g^+$-state dynamics is present, at $\lambda=1025$\,nm both signals contribute.
}
\label{fig:AllFT}
\end{center}
\end{figure}
As a consequence of the high-contrast WP oscillations that persist up to $\gtrsim 1.5$\,ns we may expect to obtain Fourier transform spectra with high resolution. Fig.~\ref{fig:AllFT} gives an overview of the Fourier spectra of the pump-probe transients recorded at selected laser wavelengths. One can clearly distinguish 2 different sets of frequencies: The spectrum at $\lambda=970$\,nm is dominated by the fundamental frequency component $\omega_a\approx 13.3$\,cm$^{-1}$ and higher harmonics up to $\omega_{a,\ell}\approx\ell\omega_a$, where $\ell=2,3,4$. As detailed in Sec.~\ref{sec:analysis}, this frequency is assigned to WP dynamics excited by resonant impulsive stimulated Raman scattering (RISRS) in the lowest triplet state $a^3\Sigma_u^+$ according to the excitation pathway sketched in Fig.~\ref{fig:Pots}~(b)~\cite{Riedle:1996}. The fundamental frequency corresponds to energy level spacings between neighboring vibrational levels, \textit{i.\,e.} $\omega_a=(E_{v''}-E_{v''-1})/(hc)$. Higher harmonics $\omega_{a,\ell}=(E_{v''}-E_{v''-\ell})/(hc)$, where $\ell=2,\,3,\,4$, are present in the spectrum as a consequence of overtone transitions $v''=\ell\leftarrow v''=0$ which are allowed due to the anharmonic shape of the potential $V_a$. However, their amplitudes are strongly suppressed compared to the one of $\omega_a$.

\begin{table}
 \centering
\begin{tabular}{|c|c|c|c|c|}
  \hline
   & \multicolumn{2}{|c|}{$^{85}$Rb$_2$} & \multicolumn{2}{|c|}{$^{87}$Rb$_2$}\\
   \hline
  $v''$ & $\omega_{v''}$ & $B_{v''}$ & $\omega_{v''}$ & $B_{v''}$ \\
  \hline
  0 & 0 & 0.01095 & 0 & 0.0107 \\
  1 & 13.321 & 0.01083 & 13.170 & 0.01054 \\
  2 & 26.263 & 0.01065 & 25.972 & 0.01037 \\
  3 & 38.827 & 0.01047 & 38.402 & 0.01020 \\
  4 & 51.016 & 0.01030 & 50.478 & 0.01003 \\
  5 & 62.819 & 0.01015 & 62.122 & 0.00985 \\
  \hline
\end{tabular}
 \caption{Experimental values for the lowest vibrational frequencies $\omega_{v''}$ of the $a$-state of Rb$_2$ with respect to $\omega_{v''=0}$. The corresponding rotational constants $B_{v''}$ are calculated from the fitted Morse potential. Units are cm$^{-1}$.}
  \label{tab:levelsa}
\end{table}

\begin{figure}[t!]
\begin{center}
{
\includegraphics[width=0.51\textwidth]{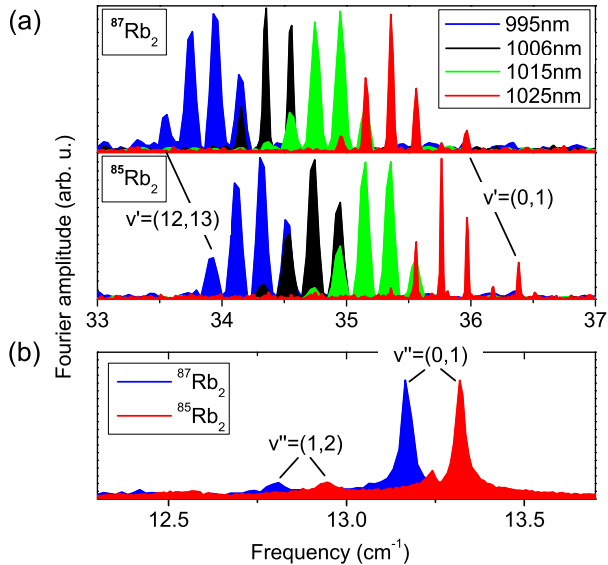}}
\caption{Detailed views of the Fourier spectra around the fundamental frequencies of wave packet dynamics in the first excited triplet state $(1)^3\Sigma_g^+$ (a) and in the lowest triplet state $a^3\Sigma_u^+$ (b). In panels (a), different colors represent measurements at different laser wave lengths. The vibrational frequencies of the isotopologues $^{85}$Rb$_2$ and $^{87}$Rb$_2$ are considerably shifted.
}
\label{fig:FTzoom}
\end{center}
\end{figure}

Fig.~\ref{fig:FTzoom} (b) shows a closeup view of the spectrum in the range of the fundamental frequency $\omega_a=(E_1-E_0)/(hc)$ recorded at laser wavelength 1025\,nm. The high spectral resolution allows to separate the fundamental frequencies $\omega_a\approx 13.2$ and $13.3$\,cm$^{-1}$ of the two isotopologues $^{85}$Rb$_2$ and $^{87}$Rb$_2$. The small peak in between at about $13.25$\,cm$^{-1}$ comes from the mixed isotopologue $^{85}$Rb$^{87}$Rb which adds to the $^{85}$Rb$_2$ signal as a consequence of insufficient mass resolution of the QMS detector. In addition, small peaks shifted to lower frequencies by about $0.4$\,cm$^{-1}$ are visible in the spectrum as beat frequencies between excited vibrational states, $(E_2-E_1)/(hc)$. Their low peak intensity directly reflects the weak thermal population of excited vibrational states $v''>0$ due to the low temperature of the molecules formed on He nanodroplets. The peak positions $\omega_{v''}$ for $1\leq v''\leq 5$ with respect to $\omega_0$ are determined by fitting Lorentzian model curves to the data and are summarized in Table~\ref{tab:levelsa}. The uncertainty associated with each value of $\omega_{v''}$ is estimated from the precision of the fit results to be $\lesssim 0.003$\,cm$^{-1}$. In addition to the level frequencies, Table~\ref{tab:levelsa} lists the rotational constants calculated using the Morse potential extracted from the vibrational spectrum (cf. Sec.~\ref{sec:analysis}).

It is important to note the high spectral resolution obtained using these fs pump-probe measurements. \textit{E.\,g.}, the peaks in Fig.~\ref{fig:FTzoom} (b) have full widths at half maximum (FWHM) $\Delta\omega_a^{FWHM}\approx 0.03\,$cm$^{-1}$ ($\approx 900\,$MHz). In particular, the resolution is higher or comparable with rotational, fine structure, and hyperfine splittings in the $a$-state. However, besides the vibrational spacings no additional beat frequencies are observed. The fact that no rovibrational cross-beatings of the type $\omega_{rovib}=(E_{v',J'=J''+1}-E_{v'-1,J'=J''-1})/(hc)$ are observed may have two reasons. Since beatings between rovibrational states with equal $J'$, $\omega_{vib}=(E_{v',J'=J''\pm 1}-E_{v'-1,J'=J''\pm 1})/(hc)$ are statistically favored with respect to the cross-beatings $\omega_{rovib}$, summing over all frequency contributions (assuming thermally populated $J'$-states) effectively reproduces the pure vibrational spectrum on top of a broad background. Additionally, the initial coupling to the He droplet prior to desorption may induce rotational decoherence and thereby quench the rovibrational beating signals. The fact that the pure rotational dynamics with frequencies $\omega_{rot}=(E_{v',J'=J''+1}-E_{v',J'=J''-1})/(hc)=2B_v(J'+1)$ is completely missing in all our measured transients for delay times $\gtrsim 10$\,ps supports the latter interpretation.

In both cases, the contribution of different rotational states to the signal only becomes observable indirectly through the rotation-vibration coupling, which leads to slightly different vibrational frequencies for different $J'$-states. Summing over the $J'$-state contributions then leads to a slight broadening and shifting of the band origins by approximately $\Delta\omega\approx -\alpha_e k_{\mathrm{B}}T/(B_vhc)$, where $\alpha_e$ denotes the rotation-vibration coupling constant and $T$ stands for the rotational temperature~\cite{Gruebele:1992}. From \textit{ab initio} potentials (Sec.~\ref{sec:analysis}) we find the value $\alpha_e^a \approx 1.6\cdot 10^{-4}$\,cm$^{-1}$, which would lead to a systematic shift of the determined vibrational frequencies by $\Delta\omega_a\approx -0.004\,$cm$^{-1}$ when assuming full thermalization to the He droplet temperature of $T=0.4\,$K.



At laser wavelength $\lambda=1006$\,nm, the Fourier spectrum in Fig.~\ref{fig:AllFT} (center panel) shows the fundamental frequency $\omega_{\Sigma g}=(E_{v'}-E_{v'-1})/(hc)\approx 35$\,cm$^{-1}$ and higher harmonics $\omega_{\Sigma g,\ell}=(E_{v'}-E_{v'-\ell})/(hc)$ up to $\ell=3$. These frequencies reflect the wave packet propagation in the first excited triplet state of Rb$_2$, $(1)^3\Sigma_g^+$, excited according to the scheme shown in Fig.~\ref{fig:Pots}~(a). At $\lambda=1025$\,nm, both sets of frequencies are present with roughly equal intensities.

The detailed view of the Fourier spectrum around $\omega_{\Sigma g}$, shown in Fig.~\ref{fig:FTzoom} (a), reveals progressions of vibrational frequency differences for the two isotopologues $^{85}$Rb$_2$ and $^{87}$Rb$_2$. Different peak colors indicate measurements using selected laser wavelengths leading to the excitation of different sets of vibrational states $v'$ according to energy conservation and Franck-Condon principle. All individual spectra are normalized to the height of the dominant peak. From these Fourier spectra frequency differences $\omega_{v'}=(E_{v'}-E_{v'-1})/(hc)$ between neighboring levels $v'=(0,1)$ up to $v'=(12,13)$ can be determined with high precision. Table~\ref{tab:levelsb} summarizes the values of $\omega_{v'}$ obtained from Lorentzian fits with an uncertainty $\lesssim 0.005$\,cm$^{-1}$. In addition, rotational constants calculated from Morse potentials obtained from the measured data are listed (Sec.~\ref{sec:analysis}). The systematic shift due to rotation-vibration coupling is estimated to $\Delta\omega_{\Sigma g}\approx -0.0015\,$cm$^{-1}$, given the rotation-vibration coupling constant $\alpha_e^{\Sigma g}\approx 5.8\cdot 10^{-5}$\,cm$^{-1}$ and the rotational temperature $T=0.4$\,K.
\begin{table}
 \centering
\begin{tabular}{|c|c|c|c|c|}
  \hline
   & \multicolumn{2}{|c|}{$^{85}$Rb$_2$} & \multicolumn{2}{|c|}{$^{87}$Rb$_2$}\\
   \hline
  $v'$ & $\Delta\omega_{v'}$ & $B_{v'}$ & $\Delta\omega_{v'}$ & $B_{v'}$ \\
  \hline
  0 & & 0.01486 & & 0.01452\\
  (0,1) & 36.382  & 0.0148 & 35.966 & 0.01446 \\
  (1,2) & 36.179 & 0.01474 & 35.765 & 0.0144 \\
  (2,3) & 35.972 & 0.01468 & 35.565 & 0.01434 \\
  (3,4) & 35.763 & 0.01462 & 35.361 & 0.01429 \\
  (4,5) & 35.555 & 0.01456 & 35.156 & 0.01423 \\
  (5,6) & 35.345 & 0.0145 & 34.945 & 0.01417 \\
  (6,7) & 35.162 & 0.01444 & 34.764 & 0.01411 \\
  (7,8) & 34.953 & 0.01438 & 34.558 & 0.01405 \\
  (8,9) & 34.737 & 0.01432 & 34.350 & 0.01400 \\
  (9,10) & 34.523 & 0.01426 & 34.139 & 0.01394 \\
  (10,11) & 34.332 & 0.0142 & 33.957 & 0.01388 \\
  (11,12) & 34.119 & 0.01414 & 33.745 & 0.01382 \\
  (12,13) & 33.910 & 0.01408 & 33.533 & 0.01376 \\
  (13,14) & 33.714 & 0.01402 & 33.331 & 0.01370 \\
  \hline
\end{tabular}
 \caption{Experimental level spacings $\omega_{v'}=(E_{v'}-E_{v'-1})/(hc)$ and calculated rotational constants $B_{v'}$ of the lowest vibrations $v'$ of the first excited state $(1)^3\Sigma_g^+$ of Rb$_2$.}
  \label{tab:levelsb}
\end{table}

\begin{figure}[t!]
\begin{center}
{
\includegraphics[width=0.51\textwidth]{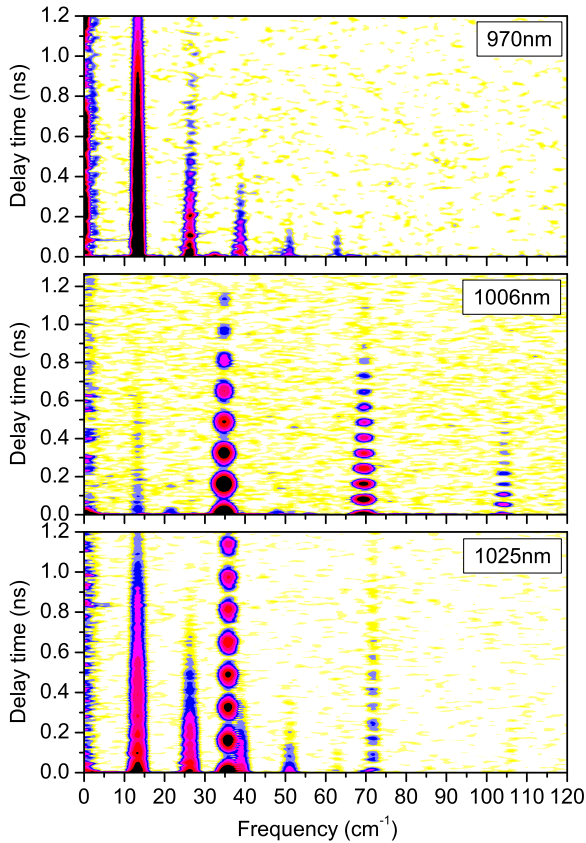}}
\caption{Spectrogram representation of the vibrational wave packet dynamics of Rb$_2$ at selected laser wavelengths. Continuous vertical bands correspond to monotonous oscillation amplitudes, dotted bands reflect revivals of the wave packet motion.
}
\label{fig:spectrograms}
\end{center}
\end{figure}
The temporal behavior of individual spectral components can be visualized by means of the spectrogram representation~\cite{Rutz:1997}. Fig.~\ref{fig:spectrograms} shows the same data as Fig.~\ref{fig:AllFT}, analyzed by Fourier transformation inside a 10 ps-time window of Gaussian shape which slides across the entire data set. The vertical dark lines in the top panel represent constant frequencies $\omega_{a,\ell}$ as a function of delay time. The fading away of signal intensity at long delay times $\geq 100$\,ps reflects damping of the WP motion, as mentioned above. Clearly, amplitudes of the higher harmonics $\omega_{a,\ell}$, where $\ell >1$, fall off faster as compared to the fundamental frequency component $\omega_{a}$. At $\lambda=1006$\,nm, full and fractional revivals of the wave packet dynamics nicely show up as periodic intensity maxima as a function of delay time. In this representation, half-period and one-third-period revivals occurring at $t_{1/2}=k\cdot T_{rev}/4$ and at $t_{1/3}=k\cdot T_{rev}/6$, where $k=1,2,3,\dots$, respectively, are clearly discerned as second and third harmonic beats (see Sec.~\ref{sec:intro}). Again, WP coherence fades away increasingly quickly for the higher harmonics $\omega_{\Sigma g,\ell}$.

In contrast to the pronounced revival structure of the $(1)^3\Sigma_g^+$-state dynamics, the $a$-state dynamics features only very weak revivals at integer multiples of about 80\,ps (Fig.~\ref{fig:ppscan}). This is due to the different distributions of population in vibrational level manifolds $v''$ and $v'$. Since in the $a$-state predominantly $v''=0$ is populated by the pump pulse and to a much lesser extent $v''=1$ and higher $v''$-states, the dynamics primarily results from the beating between the two levels $v''=0$ and $v''=1$. This causes WP oscillations with only weak revival structure. In contrast, in the $(1)^3\Sigma_g^+$-state a number of $v'$-levels are roughly equally populated which gives rise to high contrast revival features.

It is interesting to note that even when the pump-probe-transient is entirely dominated by WP-dynamics in the $a$-state (cf. Fig.~\ref{fig:spectrograms} (a)), there are still contributions from $(1)^3\Sigma_g^+$-state dynamics that rapidly disappear at delay times $t\gtrsim 10\,$ps (signal at $\omega_{\Sigma g}\approx 33\,$cm$^{-1}$). We take this as an indication that, while WP-dynamics at $t\gtrsim 10\,$ps reflects unperturbed gas-phase vibrations, frequencies and amplitudes of WP-oscillations at short delay times are affected by He matrix interactions. In particular, the unexpectedly prominent $a$-state dynamics associated with RISRS-excitation may be due to strong quenching and decoherence of vibrations in the excited $(1)^3\Sigma_g^+$-state by the He droplet as compared to damping of $a$-state vibrations. The effect of the He environment on the rovibrational dynamics of different ground and excited states of Rb$_2$ and other molecules will be further studied both experimentally and theoretically.

\section{Analysis of Fourier spectra}
\label{sec:analysis}
\begin{figure}[t!]
\begin{center}
{
\includegraphics[width=0.51\textwidth]{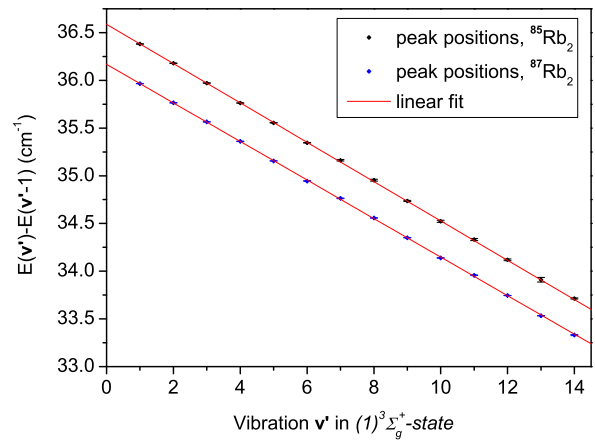}}
\caption{Linear fits of the experimental frequency differences between adjacent vibrational levels of the $(1)^3\Sigma_g^+$-state of $^{85}$Rb$_2$ and $^{87}$Rb$_2$.
}
\label{fig:BirgeSponer}
\end{center}
\end{figure}
\begin{figure}[t!]
\begin{center}
{
\includegraphics[width=0.51\textwidth]{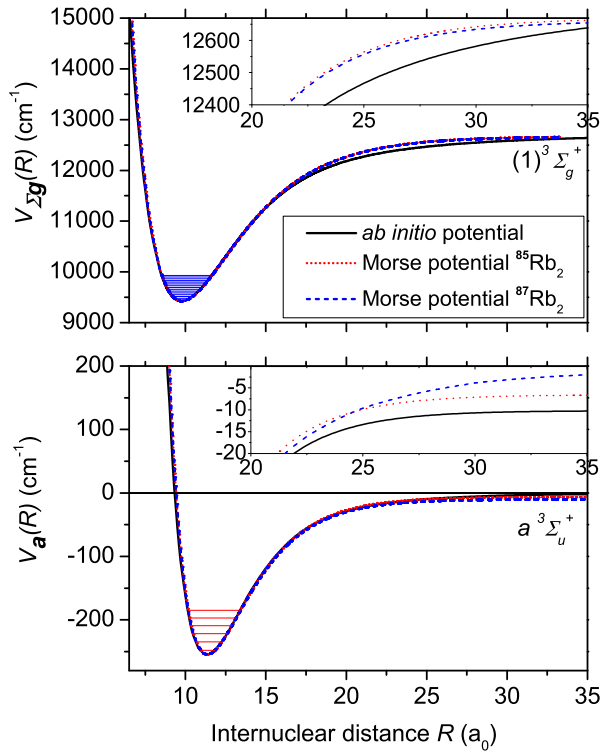}}
\caption{Morse potentials of the first excited triplet state $(1)^3\Sigma_g^+$ (a) and of the lowest triplet state $a^3\Sigma_u^+$ (b) derived from the measured vibrational frequencies compared to \textit{ab initio} calculations. The insets show magnified views of the long-range parts of the potentials.
}
\label{fig:Morse}
\end{center}
\end{figure}

In order to compare the experimental results with theoretical predictions, the experimental data are fitted to the energy terms of the Morse potential (see Sec.~\ref{sec:intro}). Thus, by determining $\omega$ and $E_D$ from the experimental data a Morse potential can be specified as approximation to the real molecular potential curve. Only the position of the potential well, $R_e$, remains undefined and will be adjusted by comparing with \textit{ab initio} potentials published as supplementary material in Ref.~\cite{Lozeille:2006}. Energy differences between neighboring levels are given by the linear dependence $\Delta E=E_{v}-E_{v-1}=\hbar\omega (1-v\,\omega /(2E_D))$. The linear fits of the experimental values of $\omega_{\Sigma g}$ for the $(1)^3\Sigma_g^+$-state of the two homonuclear isotopologues of Rb$_2$ are shown in the Birge-Sponer plot (Fig.~\ref{fig:BirgeSponer}). The error bars are barely visible on the given frequency scale. The agreement between the experimental data and the linear model is excellent in the energy range of $E_{v'}$ for $0\leq v'=0\leq 13$. The same analysis is performed on the data for the $a$-state. The resulting Morse potential parameters are summarized in Table~\ref{tab:Morse}. The associated potential curves for the $(1)^3\Sigma_g^+$ and $a$-states of $^{85}$Rb$_2$ and $^{87}$Rb$_2$ are plotted in Fig.~\ref{fig:Morse} (a) and (b) as dashed and dotted lines, respectively. The good agreement of the curves for the two isotopologues highlights the high level of precision of the determination of vibrational frequencies from our data.

\begin{table}
 \centering
\begin{tabular}{|c|c|c|c|c|c|}
  \hline
   State & Rb$_2$ & $E_D/$cm$^{-1}$ & $\Delta E_D/$cm$^{-1}$ & $a/$\AA$^{-1}$ & $\Delta a/$\AA$^{-1}$\\
   \hline
  $a$ & 85 & 247.90 & 0.40 & 0.69044 & 0.0006 \\
      & 87 & 244.3 & 3.2 & 0.6963 & 0.005 \\
  \hline
  $(1)^3\Sigma_g^+$ &  85 & 3246.42 & 4.3 & 0.50954 & 0.0003 \\
      &  87 & 3238.015 & 0.94 & 0.51023 & 0.00007 \\
  \hline
\end{tabular}
 \caption{Morse potential parameters and absolute uncertainties obtained from fitting the experimental frequencies by the Morse term energies.}
  \label{tab:Morse}
\end{table}
In addition to the Morse potentials, Fig.~\ref{fig:Morse} shows the \textit{ab initio} potential curves as black solid lines. As recommended in Ref.~\cite{Lozeille:2006}, a short-range effective potential has to be added to the computed potential curves. Therefore, the \textit{ab initio} $(1)^3\Sigma_g^+$-state potential is modified by adding an additional repulsive potential $V_{sr}(R)=1.613181\mathrm{e}^{-3.5R}R^{8.5}$ according to Ref.~\cite{Spiegelmann:1989}. In the same way the $a$-state potential curve is corrected by adding a short-range potential, which is tabulated in Ref.~\cite{Jeung:1997}. As expected, close to the dissociation limit the Morse potentials significantly deviate from the shape of the \textit{ab initio} potential. However, the depths of the potential wells $E_D$ of both $a$- and $(1)^3\Sigma_g^+$-states are in remarkably good agreement with the ones of the \textit{ab initio} potentials, $E_{D,\mathit{ab\,initio}}^{b}/(hc)=3313.7$\,cm$^{-1}$ and $E_{D,\mathit{ab\,initio}}^{a}/(hc)=254.9$\,cm$^{-1}$, considering the limited energy range covered by the observed vibrational levels (straight horizontal lines in Fig.~\ref{fig:Morse}).

The fact that in our experiment the revival structure of the $(1)^3\Sigma_g^+$-dynamics is very pronounced and long-lived is the consequence of the good agreement of the $(1)^3\Sigma_g^+$-state potential curve $V_{\Sigma g}$ with the Morse potential. Deviations from the shape of the Morse potential would cause the revival structure itself to dephase and the WP would only be fully recovered at much longer times. This effect, however, is not at the origin of the observed exponential damping of the oscillation amplitudes.

Using the fit parameters we can now determine the delay time when the first full revival is expected to occur, $t=T_{rev}/2=\pi/(\omega x_e)=4\pi E_D/(\hbar\omega^2)$. For $^{85}$Rb$_2$ [$^{87}$Rb$_2$] in the $(1)^3\Sigma_g^+$-state we find the values $T_{rev}/2=$161.8\,ps [165.1\,ps], which are in excellent agreement with the observation. For the $a$-state dynamics we calculate revival times $T_{rev}/2=$90.6\,ps [90.0\,ps], which approximately match the observed values.

\begin{figure}[t!]
\begin{center}
{
\includegraphics[width=0.51\textwidth]{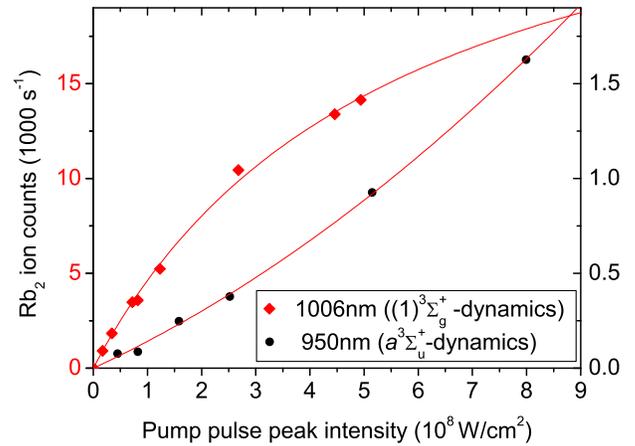}}
\caption{Dependence on laser intensity of the first exciting pump pulse of the fringe contrast of $a$ and $(1)^3\Sigma_g^+$-state dynamics.
}
\label{fig:Pdep}
\end{center}
\end{figure}
Assigning the two sets of frequencies to WP dynamics in the lowest and first excited states is further confirmed by selectively measuring the amplitude of WP dynamics in $a$ and $(1)^3\Sigma_g^+$-states as a function of varying laser intensity of the first exciting pump pulse, $I_{\mathrm{pump}}$. To selectively attenuate the pump pulse a variable attenuator is introduced in one of the two arms of the Mach-Zehnder interferometer. Great care is taken to realign pump and probe laser beams coaxially for every setting of the attenuator. The $a$-state dynamics is selectively probed by measuring the maximum pump-probe signal at laser wavelength $\lambda=950$\,nm, at which the dynamics is nearly entirely determined by WP propagation in the $a$-state. The delay time $t=80$\,ps is chosen such that transient effects by the interaction of Rb$_2$ with the He environment during desorption can be safely neglected. The $(1)^3\Sigma_g^+$-state dynamics is probed at $\lambda=1006$\,nm at the maximum of the first full revival ($t=160$\,ps). In order to isolate the pure effect of the pump pulse in the pump-probe scheme the background count rates resulting from direct 3PI when applying only the pump pulse or only the probe pule are subtracted from the data. The resulting state selective signals are shown as symbols in Fig.~\ref{fig:Pdep}. The solid lines represent fit curves to the experimental data to guide the eye.
The $\omega_a$-data are modeled by a second order polynomial to account for the second order process of the RISRS scheme. The $\omega_{\Sigma g}$-data are modeled by a simple saturation model to account for saturation of the strong resonant transition $(1)^3\Sigma_g^+\leftarrow a$. Clearly, the data follow qualitatively different behavior which confirms the excitation schemes shown in Fig.~\ref{fig:Pots} and thus supports the assignment of the observed frequency components presented in Fig.~\ref{fig:AllFT} to $a$- and $(1)^3\Sigma_g^+$-state vibrations.

\begin{figure}[t!]
\begin{center}
{
\includegraphics[width=0.45\textwidth]{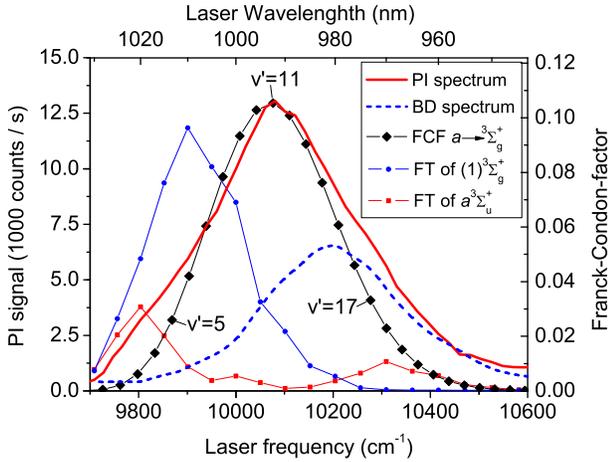}}
\caption{Comparison of the calculated Franck-Condon factors (FCFs) for the transition $(1)^3\Sigma_g^+(v')\leftarrow a(v''=0)$ with the measured photoionization (PI) spectrum, with the beam depletion (BD) spectrum and with the amplitudes of the Fourier transform (FT) frequency components of $a$ and $(1)^3\Sigma_g^+$-state wave packet dynamics.
}
\label{fig:dipole}
\end{center}
\end{figure}
The correct identification of the $(1)^3\Sigma_g^+$-state WP dynamics is further confirmed by comparing the one-pulse PI-spectrum shown in Fig.~\ref{fig:PIspectra} in the wavelength range around 1000\,nm with calculated Franck-Condon factors (FCFs) for the transition $(1)^3\Sigma_g^+(v')\leftarrow a(v''=0)$. The FCFs are computed using LeRoy's LEVEL 8.0 program~\cite{level} with the \textit{ab initio} potential as input. The resulting values are plotted in Fig.~\ref{fig:dipole} as black diamonds, the experimental spectrum is shown as red solid line. The agreement between measurement and calculation regarding peak positions and widths is quite good, in particular when considering additional broadening of the fs PI spectrum due to the spectral bandwidth of the fs-laser as well as saturation effects. However, it has to be assumed that this agreement relies on the accidental cancelation of He-matrix induced frequency shifts and the effect of near-resonant higher lying electronic states of Rb$_2$, since the peak measured by beam depletion (BD) at lower laser intensity (blue dotted line) is blue shifted by about 100\,cm$^{-1}$ with respect to the PI maximum. Thus, the discrepancy between the line positions in the PI and BD spectra is attributed to enhanced ionization probability due to higher lying resonances at certain laser wavelengths.

The additional red squares and blue circles represent amplitudes of the Fourier components, $A_a$ and $A_{\Sigma g}$, of the fundamental frequencies $\omega_a$ and $\omega_{\Sigma g}$, respectively. The data have been corrected for varying laser intensity when scanning the laser wavelength. Interestingly, the $A_{\Sigma g}$ amplitude spectrum is shifted to smaller frequencies by about 200\,cm$^{-1}$ with respect to the one-pulse PI spectrum. This is an effect of the additional frequency selectivity imposed by the transition from the $(1)^3\Sigma_g^+$-state into the ionic continuum of Rb$_2$. In fact, when incorporating into the calculation the FCFs for the transition from the $(1)^3\Sigma_g^+$-state to the intermediate near-resonant state $4^3\Sigma_u^+$, good agreement with the $A_{\Sigma g}$-amplitude spectrum (blue circles) is achieved.

The $A_a$ amplitude spectrum features two maxima in the wings of the one-pulse PI spectrum around $\lambda=1020$\,nm and $\lambda=970$\,nm, and a minimum at the peak position of the PI spectrum around  $\lambda=990$\,nm. Again, this is attributed to the influence of various near-resonant higher-lying states on the transition to the ionic state driven by the probe pulse. This interpretation is backed by the analysis of the initial phases of the $a$-state oscillations obtained from band-pass filtered pump-probe transients. At $\lambda=1020$\,nm, the initial phase $\phi_{0,\, 1020\,\mathrm{nm}}=0.26\cdot 2\pi$ significantly differs from the one obtained at $\lambda=970$\,nm, $\phi_{0,\,970\,\mathrm{nm}}=0.68\cdot 2\pi$, which highlights the presence of two distinct detection windows.


\section{Excitation around 850\,\lowercase{nm}}
\label{sec:singlet}
\begin{figure}[t!]
\begin{center}
{
\includegraphics[width=0.51\textwidth]{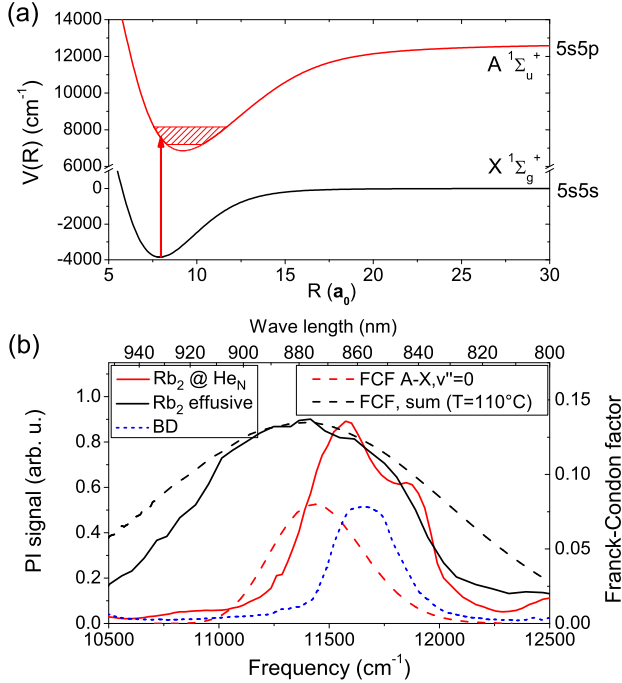}}
\caption{(a) Potential energy curves of ground and first excited singlet states of Rb$_2$. (b) Photoionization spectra of Rb$_2$ in an effusive beam and of Rb$_2$ attached to He nanodroplets (solid lines) in comparison with the beam depletion spectrum (blue dotted line) and with calculated Franck-Condon factors (FCFs) for the transition $A^1\Sigma_u^+\leftarrow X^1\Sigma_g^+$ (dashed lines).
}
\label{fig:singlet}
\end{center}
\end{figure}
The signal of Rb$_2^+$ photo-ions measured on He nanodroplets features an additional maximum around 850\,nm, which roughly coincides with a broad feature measured in the effusive beam (Fig.~\ref{fig:PIspectra} (b)). Since only strongly bound singlet molecules are expected to be formed in a hot effusive beam, these peaks are assigned to singlet transitions. This assignment is further supported by the comparison with calculated FCFs, as shown in Fig.~\ref{fig:singlet}. The red dashed line represents the Frank-Condon factors for the transitions $A^1\Sigma_u^+\leftarrow X^1\Sigma_g^+(v''=0)$ from the lowest vibrational state $v''=0$ of the singlet ground state $X$ into vibrational states of the first excited singlet state $A$. This distribution is in reasonable agreement with the left peak of the droplet spectrum of Rb$_2$. The slight deviation of peak positions may be due to perturbations by the He matrix or due to the influence of higher lying resonances in the 3PI process. In addition, spin-orbit coupling between $A$ and $b^3\Pi_u$-states are known to induce strong perturbations~\cite{Beuc:2007}. The black dashed line is obtained by summing over Franck-Condon spectra assuming thermally populated $v''$-levels at the temperature of the Rb vapor cell (380\,K). The resulting profile matches the measured PI spectrum of effusive Rb$_2$ quite well. The peak in the BD spectrum (blue dotted line) is narrower but centered at the same position with respect to the PI signal. The frequency shift of about 300\,cm$^{-1}$ to the blue as compared to the calculated Franck-Condon profile is attributed to the effect of the He matrix.

Besides the mentioned singlet transition, this part of the He droplet spectrum measured at the mass of Rb$_2$ may also result from fragmentation of Rb$_3^+\rightarrow$Rb$_2+$Rb and possibly of larger oligomers following PI, as suggested by the resemblance of Rb$_2$ and Rb$_3$ spectra (Fig.~\ref{fig:PIspectra} (b) and (c)). In particular, at about 840\,nm both Rb$_2$ and Rb$_3$ droplet spectra feature coinciding local maxima.  Unfortunately, no WP oscillations in the pump-probe transients of Rb$_2$ are observed in this wavelength range. However, around 850\,nm, in the region of the transition $(2)^4E'\leftarrow (1)^4A_2'$ of Rb$_3$ in the quartet manifold, we do observe WP oscillations with low contrast when setting the ion detector to the mass of Rb$_3$. These data will be presented in a forthcoming publication. The fact that these oscillations are not observed on the Rb$_2$-mass indicates that the contribution of Rb$_3$ fragmentation to the Rb$_2$-signal is smaller than about 20\,$\%$ of the signal intensity.

\section{Excitation around 750\,\lowercase{nm}}
\label{sec:rightpeak}
\begin{figure}[t!]
\begin{center}
{
\includegraphics[width=0.51\textwidth]{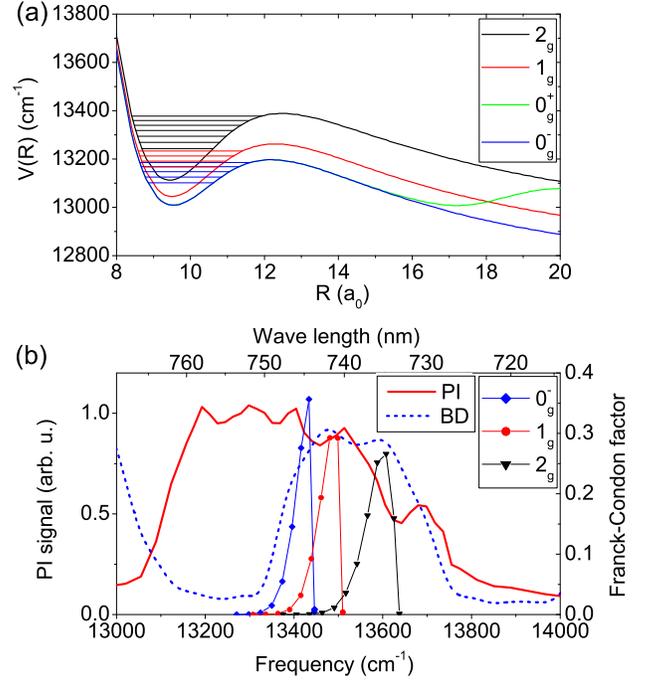}}
\caption{Comparison of calculated Franck-Condon factors for the transition $(1)^3\Pi_g\leftarrow a(v''=0)$ with the measured photoionization and with beam depletion spectra.
}
\label{fig:rightpeak}
\end{center}
\end{figure}
The Rb$_2$ PI spectrum displays an additional, slightly structured maximum around 750\,nm, shown in Fig.~\ref{fig:rightpeak} (b) as red solid line. The BD spectrum (blue dotted line) in this spectral range significantly differs from the PI spectrum by featuring a much narrower peak. The lack of effusive signal at that wavelength already points at transitions in triplet molecules. Indeed, the transition from the lowest triplet state $a$ to the second excited state $(1)^3\Pi_g$ is found to fall into this wavelength region. Since the $(1)^3\Pi_g$ state is significantly perturbed by first-order spin-orbit coupling, fine-structure states $0_g^-$, $1_g$, and $2_g$ have to be considered. The corresponding potential curves are depicted in Fig.~\ref{fig:rightpeak} (a) including the vibrational levels probed in the experiment. The FCFs for transitions from $a(v''=0)$ into these states are shown in Fig.~\ref{fig:rightpeak} (b) in comparison with the measured PI and BD spectra. While the BD spectrum agrees quite well with the Franck-Condon spectrum when accounting for the laser bandwidth, both width and fine-structure of the PI spectrum are not correctly reproduced. The fact that the line intensities of the low-frequency part of the spectrum ($\lesssim 13400$\,cm$^{-1}$) varies much stronger as a function of laser intensity than the high-frequency part indicates that near-resonant transitions at higher excitations of Rb$_2$ perturb the spectrum particularly strongly in the vicinity of the $(1)^3\Pi_g\leftarrow a$-transition.

This transition has been studied before using cw excitation and detection of laser-induced fluorescence as well as magnetic circular dichroism~\cite{Auboeck:2007}. The cw excitation spectrum featured 2 broad maxima around 735\,nm and 744\,nm, quite similar to our BD spectrum. The interpretation by Aub\"ock \textit{et al.} was based on a molecular Hamiltonian containing spin-orbit coupling and a ``crystal field'' interaction with the surface of the He droplet.

\begin{figure}[t!]
\begin{center}
{
\includegraphics[width=0.51\textwidth]{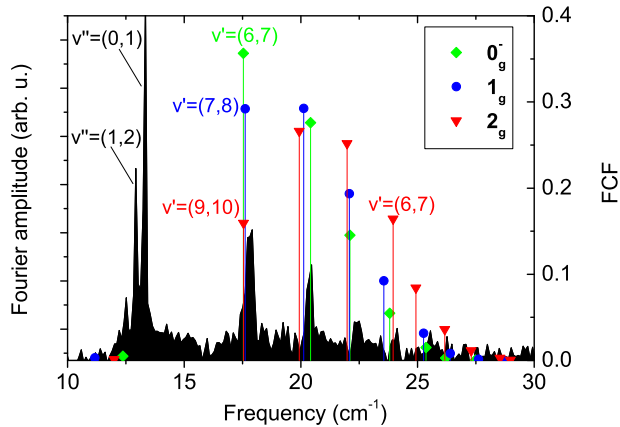}}
\caption{Fourier spectrum of the wave packet dynamics in spin-orbit states of $(1)^3\Pi_g$.
}
\label{fig:rightpeakFT}
\end{center}
\end{figure}
Pump-probe transients around 750\,nm are modulated by WP oscillations, though the fringe contrast is much lower than the one observed around 1000\,nm. The Fourier spectrum of the delay scan recorded at $\lambda =750$\,nm is shown in Fig.~\ref{fig:rightpeakFT}. The strongest frequency components are the fundamental ground state frequency $\omega_a (0,1)\approx 13.3\,$cm$^{-1}$ as well as $\omega_a (1,2)\approx 12.9$\,cm$^{-1}$. Thus, the dominant excitation mechanism is RISRS. In addition, frequency components show up at 17.8\,cm$^{-1}$, 20.4\,cm$^{-1}$, 22.4\,cm$^{-1}$, 24.1\,cm$^{-1}$, and at 25.6\,cm$^{-1}$. This Fourier spectrum is in good agreement with the expected transition frequencies and corresponding FCFs of the transitions $2_g,\,1_g,\,0_g^-\,(v')\leftarrow a(v''=0)$, depicted as stick spectra in Fig.~\ref{fig:rightpeakFT} (b). Due to the accidental near-coincidence of vibrational level spacings in the three fine-structure states, however, it is not possible to ascertain which one predominantly determines the wave-packet dynamics.

\section{Conclusion}
\label{sec:conclusion}
In conclusion, we have systematically investigated femtosecond photo-ionization spectra of Rb$_2$ attached to He nanodroplets in the wavelength range from $\lambda=700$ to $\lambda=1030\,$nm. Three transitions between low-lying triplet states ($(1)^3\Sigma_g^+\leftarrow a$ and $(1)^3\Pi_g\leftarrow a$) and singlet states ($A\leftarrow X$) have been identified. One-color pump-probe transients around $\lambda=1000\,$nm reveal pronounced modulations due to WP oscillations in the first excited triplet state $(1)^3\Sigma_g^+$ as well as in the lowest triplet state $a^3\Sigma_u^+$. Fractional revivals of the $(1)^3\Sigma_g^+$-state dynamics are visible with high contrast and long coherence times up to 1.5\,ns due to the low internal temperature of the molecules formed on He nanodroplets. This allows to extract vibrational frequencies with high precision. The data are well fitted by Morse-like potential curves and are in good agreement with \textit{ab initio} potentials.

With this work we show that valuable spectroscopic information can be obtained from femtosecond spectroscopy of doped He nanodroplets. In the future the details of the dynamic coupling of dopant molecules with the surrounding He quantum fluid will be further analyzed. In particular, the choice of dopant molecules with different coupling strengths to the He droplet, ranging from surface-bound alkali dimers to deeply solvated molecules, \textit{e.\,g.} I$_2$, in combination with the potential to compare superfluid $^4$He and normal fluid $^3$He droplets offer great chances for studying the dynamics of complex quantum systems.


\section{Acknowledgement}
Stimulating discussions with C.P. Schulz, M. Schlesinger, W. Strunz and R. Wester
are gratefully acknowledged. This work
is supported by the Deutsche Forschungsgemeinschaft.

\bibliographystyle{apsrev}

\begin{thebibliography}{47}
\expandafter\ifx\csname natexlab\endcsname\relax\def\natexlab#1{#1}\fi
\expandafter\ifx\csname bibnamefont\endcsname\relax
  \def\bibnamefont#1{#1}\fi
\expandafter\ifx\csname bibfnamefont\endcsname\relax
  \def\bibfnamefont#1{#1}\fi
\expandafter\ifx\csname citenamefont\endcsname\relax
  \def\citenamefont#1{#1}\fi
\expandafter\ifx\csname url\endcsname\relax
  \def\url#1{\texttt{#1}}\fi
\expandafter\ifx\csname urlprefix\endcsname\relax\def\urlprefix{URL }\fi
\providecommand{\bibinfo}[2]{#2}
\providecommand{\eprint}[2][]{\url{#2}}

\bibitem[{\citenamefont{Zewail}(1994)}]{Zewail:1994}
\bibinfo{author}{\bibfnamefont{A.}~\bibnamefont{Zewail}},
  \emph{\bibinfo{title}{Femtochemistry}} (\bibinfo{publisher}{World Scientific,
  Singapore, 1994}, \bibinfo{year}{1994}).

\bibitem[{\citenamefont{Manz and W{\"o}ste}(1995)}]{Manz:1995}
\bibinfo{editor}{\bibfnamefont{J.}~\bibnamefont{Manz}} \bibnamefont{and}
  \bibinfo{editor}{\bibfnamefont{L.}~\bibnamefont{W{\"o}ste}}, eds.,
  \emph{\bibinfo{title}{Femtosecond Chemistry}} (\bibinfo{publisher}{VCH,
  Weinheim}, \bibinfo{year}{1995}).

\bibitem[{\citenamefont{Chergui}(1995)}]{Chergui:1995}
\bibinfo{editor}{\bibfnamefont{M.}~\bibnamefont{Chergui}}, ed.,
  \emph{\bibinfo{title}{Femtochemistry}} (\bibinfo{publisher}{World Scientific,
  Singapore}, \bibinfo{year}{1995}).

\bibitem[{\citenamefont{Aymar and Dulieu}(2005)}]{Aymar:2005}
\bibinfo{author}{\bibfnamefont{M.}~\bibnamefont{Aymar}} \bibnamefont{and}
  \bibinfo{author}{\bibfnamefont{O.}~\bibnamefont{Dulieu}},
  \bibinfo{journal}{J. Chem. Phys.} \textbf{\bibinfo{volume}{122}},
  \bibinfo{pages}{204302} (\bibinfo{year}{2005}).

\bibitem[{\citenamefont{Rutz et~al.}(1996)\citenamefont{Rutz, de~Vivie-Riedle,
  and Schreiber}}]{Rutz:1996}
\bibinfo{author}{\bibfnamefont{S.}~\bibnamefont{Rutz}},
  \bibinfo{author}{\bibfnamefont{R.}~\bibnamefont{de~Vivie-Riedle}},
  \bibnamefont{and}
  \bibinfo{author}{\bibfnamefont{E.}~\bibnamefont{Schreiber}},
  \bibinfo{journal}{Phys. Rev. A} \textbf{\bibinfo{volume}{54}},
  \bibinfo{pages}{306} (\bibinfo{year}{1996}).

\bibitem[{\citenamefont{Zhang et~al.}(2003)\citenamefont{Zhang, Gador, and
  Hansson}}]{Zhang:2003}
\bibinfo{author}{\bibfnamefont{B.}~\bibnamefont{Zhang}},
  \bibinfo{author}{\bibfnamefont{N.}~\bibnamefont{Gador}}, \bibnamefont{and}
  \bibinfo{author}{\bibfnamefont{T.}~\bibnamefont{Hansson}},
  \bibinfo{journal}{Phys. Rev. Lett.} \textbf{\bibinfo{volume}{91}},
  \bibinfo{pages}{173006} (\bibinfo{year}{2003}).

\bibitem[{\citenamefont{de~Vivie-Riedle
  et~al.}(1996)\citenamefont{de~Vivie-Riedle, Kobe, Manz, Meyer, Reischl, Rutz,
  Schreiber, and W{\"o}ste}}]{Riedle:1996}
\bibinfo{author}{\bibfnamefont{R.}~\bibnamefont{de~Vivie-Riedle}},
  \bibinfo{author}{\bibfnamefont{K.}~\bibnamefont{Kobe}},
  \bibinfo{author}{\bibfnamefont{J.}~\bibnamefont{Manz}},
  \bibinfo{author}{\bibfnamefont{W.}~\bibnamefont{Meyer}},
  \bibinfo{author}{\bibfnamefont{B.}~\bibnamefont{Reischl}},
  \bibinfo{author}{\bibfnamefont{S.}~\bibnamefont{Rutz}},
  \bibinfo{author}{\bibfnamefont{E.}~\bibnamefont{Schreiber}},
  \bibnamefont{and}
  \bibinfo{author}{\bibfnamefont{L.}~\bibnamefont{W{\"o}ste}},
  \bibinfo{journal}{J. Phys. Chem. A} \textbf{\bibinfo{volume}{100}},
  \bibinfo{pages}{7789} (\bibinfo{year}{1996}).

\bibitem[{\citenamefont{Nicole et~al.}(1999)\citenamefont{Nicole, Bouch\`ene,
  Meier, Magnier, Schreiber, and Girard}}]{Nicole:1999}
\bibinfo{author}{\bibfnamefont{C.}~\bibnamefont{Nicole}},
  \bibinfo{author}{\bibfnamefont{M.~A.} \bibnamefont{Bouch\`ene}},
  \bibinfo{author}{\bibfnamefont{C.}~\bibnamefont{Meier}},
  \bibinfo{author}{\bibfnamefont{S.}~\bibnamefont{Magnier}},
  \bibinfo{author}{\bibfnamefont{E.}~\bibnamefont{Schreiber}},
  \bibnamefont{and} \bibinfo{author}{\bibfnamefont{B.}~\bibnamefont{Girard}},
  \bibinfo{journal}{J. Chem. Phys.} \textbf{\bibinfo{volume}{111}},
  \bibinfo{pages}{7857} (\bibinfo{year}{1999}).

\bibitem[{\citenamefont{Rutz and Schreiber}(1997)}]{Rutz:1997}
\bibinfo{author}{\bibfnamefont{S.}~\bibnamefont{Rutz}} \bibnamefont{and}
  \bibinfo{author}{\bibfnamefont{E.}~\bibnamefont{Schreiber}},
  \bibinfo{journal}{Chem. Phys. Lett.} \textbf{\bibinfo{volume}{269}},
  \bibinfo{pages}{9} (\bibinfo{year}{1997}).

\bibitem[{\citenamefont{Vrakking et~al.}(1996)\citenamefont{Vrakking,
  Villeneuve, and Stolow}}]{Vrakking:1996}
\bibinfo{author}{\bibfnamefont{M.~J.~J.} \bibnamefont{Vrakking}},
  \bibinfo{author}{\bibfnamefont{D.~M.} \bibnamefont{Villeneuve}},
  \bibnamefont{and} \bibinfo{author}{\bibfnamefont{A.}~\bibnamefont{Stolow}},
  \bibinfo{journal}{Phys. Rev. A} \textbf{\bibinfo{volume}{54}},
  \bibinfo{pages}{R37} (\bibinfo{year}{1996}).

\bibitem[{\citenamefont{Lindinger et~al.}(2004)\citenamefont{Lindinger,
  Lupulescu, Plewicki, Vetter, Weber, Merli, and Wöste}}]{LindingerPRL:2004}
\bibinfo{author}{\bibfnamefont{A.}~\bibnamefont{Lindinger}},
  \bibinfo{author}{\bibfnamefont{C.}~\bibnamefont{Lupulescu}},
  \bibinfo{author}{\bibfnamefont{M.}~\bibnamefont{Plewicki}},
  \bibinfo{author}{\bibfnamefont{F.}~\bibnamefont{Vetter}},
  \bibinfo{author}{\bibfnamefont{S.~M.} \bibnamefont{Weber}},
  \bibinfo{author}{\bibfnamefont{A.}~\bibnamefont{Merli}}, \bibnamefont{and}
  \bibinfo{author}{\bibfnamefont{L.}~\bibnamefont{Wöste}},
  \bibinfo{journal}{Phys. Rev. Lett.} \textbf{\bibinfo{volume}{93}},
  \bibinfo{pages}{033001} (\bibinfo{year}{2004}).

\bibitem[{\citenamefont{Wollenhaupt et~al.}(2005)\citenamefont{Wollenhaupt,
  Engel, and Baumert}}]{Wollenhaupt:2005}
\bibinfo{author}{\bibfnamefont{M.}~\bibnamefont{Wollenhaupt}},
  \bibinfo{author}{\bibfnamefont{V.}~\bibnamefont{Engel}}, \bibnamefont{and}
  \bibinfo{author}{\bibfnamefont{T.}~\bibnamefont{Baumert}},
  \bibinfo{journal}{Ann. Rev. Phys. Chem.} \textbf{\bibinfo{volume}{56}},
  \bibinfo{pages}{25} (\bibinfo{year}{2005}).

\bibitem[{\citenamefont{Sch{\"a}fer-Bung
  et~al.}(2004)\citenamefont{Sch{\"a}fer-Bung, Mitri{\'c},
  Bona\v{c}i\'{c}-Kouteck\'{y}, Bartelt, Lupulescu, Lindinger, Vajda, Weber,
  and W{\"o}ste}}]{Bung:2004}
\bibinfo{author}{\bibfnamefont{B.}~\bibnamefont{Sch{\"a}fer-Bung}},
  \bibinfo{author}{\bibfnamefont{R.}~\bibnamefont{Mitri{\'c}}},
  \bibinfo{author}{\bibfnamefont{V.}~\bibnamefont{Bona\v{c}i\'{c}-Kouteck\'{y}%
}}, \bibinfo{author}{\bibfnamefont{A.}~\bibnamefont{Bartelt}},
  \bibinfo{author}{\bibfnamefont{C.}~\bibnamefont{Lupulescu}},
  \bibinfo{author}{\bibfnamefont{A.}~\bibnamefont{Lindinger}},
  \bibinfo{author}{\bibfnamefont{S.}~\bibnamefont{Vajda}},
  \bibinfo{author}{\bibfnamefont{S.~M.} \bibnamefont{Weber}}, \bibnamefont{and}
  \bibinfo{author}{\bibfnamefont{L.}~\bibnamefont{W{\"o}ste}},
  \bibinfo{journal}{J. Phys. Chem. A} \textbf{\bibinfo{volume}{108}},
  \bibinfo{pages}{4175} (\bibinfo{year}{2004}).

\bibitem[{\citenamefont{Baumert and Gerber}(2005)}]{Baumert:2005}
\bibinfo{author}{\bibfnamefont{T.}~\bibnamefont{Baumert}} \bibnamefont{and}
  \bibinfo{author}{\bibfnamefont{G.}~\bibnamefont{Gerber}}, in
  \emph{\bibinfo{booktitle}{Femtosecond Laser Spectroscopy}}, edited by
  \bibinfo{editor}{\bibfnamefont{P.}~\bibnamefont{Hannaford}}
  (\bibinfo{publisher}{Springer Verlag}, \bibinfo{year}{2005}), p.
  \bibinfo{pages}{chap. 9}.

\bibitem[{\citenamefont{Oldenbourg et~al.}(2000)\citenamefont{Oldenbourg, John,
  and Eden}}]{Oldenbourg:2000}
\bibinfo{author}{\bibfnamefont{A.~L.} \bibnamefont{Oldenbourg}},
  \bibinfo{author}{\bibfnamefont{P.~C.} \bibnamefont{John}}, \bibnamefont{and}
  \bibinfo{author}{\bibfnamefont{J.~G.} \bibnamefont{Eden}},
  \bibinfo{journal}{J. Chem. Phys.} \textbf{\bibinfo{volume}{113}},
  \bibinfo{pages}{11009} (\bibinfo{year}{2000}).

\bibitem[{\citenamefont{Gruebele and Zewail}(1992)}]{Gruebele:1992}
\bibinfo{author}{\bibfnamefont{M.}~\bibnamefont{Gruebele}} \bibnamefont{and}
  \bibinfo{author}{\bibfnamefont{A.~H.} \bibnamefont{Zewail}},
  \bibinfo{journal}{J. Phys. Chem. A} \textbf{\bibinfo{volume}{98}},
  \bibinfo{pages}{883} (\bibinfo{year}{1992}).

\bibitem[{\citenamefont{Chin et~al.}(2009)\citenamefont{Chin, Grimm, Julienne,
  and Tiesinga}}]{Chin:2009}
\bibinfo{author}{\bibfnamefont{C.}~\bibnamefont{Chin}},
  \bibinfo{author}{\bibfnamefont{R.}~\bibnamefont{Grimm}},
  \bibinfo{author}{\bibfnamefont{P.}~\bibnamefont{Julienne}}, \bibnamefont{and}
  \bibinfo{author}{\bibfnamefont{E.}~\bibnamefont{Tiesinga}},
  \bibinfo{journal}{Rev. Mod. Phys., \textit{in preparation}}
  (\bibinfo{year}{2009}), \eprint{arXiv:0812.1496}.

\bibitem[{\citenamefont{Masnou-Seeuws and Pillet}(2001)}]{Masnou:2001}
\bibinfo{author}{\bibfnamefont{F.}~\bibnamefont{Masnou-Seeuws}}
  \bibnamefont{and} \bibinfo{author}{\bibfnamefont{P.}~\bibnamefont{Pillet}},
  \bibinfo{journal}{Adv. Atomic Mol. Opt. Phys.} \textbf{\bibinfo{volume}{47}},
  \bibinfo{pages}{53} (\bibinfo{year}{2001}).

\bibitem[{\citenamefont{Jones et~al.}(2006)\citenamefont{Jones, Tiesinga, Lett,
  and Julienne}}]{Jones:2006}
\bibinfo{author}{\bibfnamefont{K.~M.} \bibnamefont{Jones}},
  \bibinfo{author}{\bibfnamefont{E.}~\bibnamefont{Tiesinga}},
  \bibinfo{author}{\bibfnamefont{P.~D.} \bibnamefont{Lett}}, \bibnamefont{and}
  \bibinfo{author}{\bibfnamefont{P.~S.} \bibnamefont{Julienne}},
  \bibinfo{journal}{Rev. Mod. Phys.} \textbf{\bibinfo{volume}{78}},
  \bibinfo{pages}{483} (\bibinfo{year}{2006}).

\bibitem[{\citenamefont{Lozeille et~al.}(2006)\citenamefont{Lozeille, Fioretti,
  Gabbanini, Huang, Pechkis, Wang, Gould, Eyler, Stwalley, Aymar
  et~al.}}]{Lozeille:2006}
\bibinfo{author}{\bibfnamefont{J.}~\bibnamefont{Lozeille}},
  \bibinfo{author}{\bibfnamefont{A.}~\bibnamefont{Fioretti}},
  \bibinfo{author}{\bibfnamefont{C.}~\bibnamefont{Gabbanini}},
  \bibinfo{author}{\bibfnamefont{Y.}~\bibnamefont{Huang}},
  \bibinfo{author}{\bibfnamefont{H.}~\bibnamefont{Pechkis}},
  \bibinfo{author}{\bibfnamefont{D.}~\bibnamefont{Wang}},
  \bibinfo{author}{\bibfnamefont{P.}~\bibnamefont{Gould}},
  \bibinfo{author}{\bibfnamefont{E.}~\bibnamefont{Eyler}},
  \bibinfo{author}{\bibfnamefont{W.}~\bibnamefont{Stwalley}},
  \bibinfo{author}{\bibfnamefont{M.}~\bibnamefont{Aymar}},
  \bibnamefont{et~al.}, \bibinfo{journal}{Eur. Phys. J. D}
  \textbf{\bibinfo{volume}{39}}, \bibinfo{pages}{261} (\bibinfo{year}{2006}).

\bibitem[{\citenamefont{Danzl et~al.}(2009)\citenamefont{Danzl, Mark, Haller,
  Gustavsson, Bouloufa, Dulieu, Ritsch, Hart, and N{\"a}gerl}}]{Danzl:2009}
\bibinfo{author}{\bibfnamefont{J.~G.} \bibnamefont{Danzl}},
  \bibinfo{author}{\bibfnamefont{M.~J.} \bibnamefont{Mark}},
  \bibinfo{author}{\bibfnamefont{E.}~\bibnamefont{Haller}},
  \bibinfo{author}{\bibfnamefont{M.}~\bibnamefont{Gustavsson}},
  \bibinfo{author}{\bibfnamefont{N.}~\bibnamefont{Bouloufa}},
  \bibinfo{author}{\bibfnamefont{O.}~\bibnamefont{Dulieu}},
  \bibinfo{author}{\bibfnamefont{H.}~\bibnamefont{Ritsch}},
  \bibinfo{author}{\bibfnamefont{R.}~\bibnamefont{Hart}}, \bibnamefont{and}
  \bibinfo{author}{\bibfnamefont{H.-C.} \bibnamefont{N{\"a}gerl}},
  \bibinfo{journal}{New J. Phys.} \textbf{\bibinfo{volume}{11}},
  \bibinfo{pages}{055036} (\bibinfo{year}{2009}).

\bibitem[{\citenamefont{Claas et~al.}(2006)\citenamefont{Claas, Droppelmann,
  Schulz, Mudrich, and Stienkemeier}}]{Claas:2006}
\bibinfo{author}{\bibfnamefont{P.}~\bibnamefont{Claas}},
  \bibinfo{author}{\bibfnamefont{G.}~\bibnamefont{Droppelmann}},
  \bibinfo{author}{\bibfnamefont{C.~P.} \bibnamefont{Schulz}},
  \bibinfo{author}{\bibfnamefont{M.}~\bibnamefont{Mudrich}}, \bibnamefont{and}
  \bibinfo{author}{\bibfnamefont{F.}~\bibnamefont{Stienkemeier}},
  \bibinfo{journal}{J. Phys. B} \textbf{\bibinfo{volume}{39}},
  \bibinfo{pages}{S1151} (\bibinfo{year}{2006}).

\bibitem[{\citenamefont{Claas et~al.}(2007)\citenamefont{Claas, Droppelmann,
  Schulz, Mudrich, and Stienkemeier}}]{Claas:2007}
\bibinfo{author}{\bibfnamefont{P.}~\bibnamefont{Claas}},
  \bibinfo{author}{\bibfnamefont{G.}~\bibnamefont{Droppelmann}},
  \bibinfo{author}{\bibfnamefont{C.~P.} \bibnamefont{Schulz}},
  \bibinfo{author}{\bibfnamefont{M.}~\bibnamefont{Mudrich}}, \bibnamefont{and}
  \bibinfo{author}{\bibfnamefont{F.}~\bibnamefont{Stienkemeier}},
  \bibinfo{journal}{J. Phys. Chem. A} \textbf{\bibinfo{volume}{111}},
  \bibinfo{pages}{7537} (\bibinfo{year}{2007}).

\bibitem[{\citenamefont{Toennies and Vilesov}(2004)}]{Toennies:2004}
\bibinfo{author}{\bibfnamefont{J.~P.} \bibnamefont{Toennies}} \bibnamefont{and}
  \bibinfo{author}{\bibfnamefont{A.~F.} \bibnamefont{Vilesov}},
  \bibinfo{journal}{Angew. Chem. Int. Ed.} \textbf{\bibinfo{volume}{43}},
  \bibinfo{pages}{2622} (\bibinfo{year}{2004}).

\bibitem[{\citenamefont{Stienkemeier and Lehmann}(2006)}]{Stienkemeier:2006}
\bibinfo{author}{\bibfnamefont{F.}~\bibnamefont{Stienkemeier}}
  \bibnamefont{and} \bibinfo{author}{\bibfnamefont{K.}~\bibnamefont{Lehmann}},
  \bibinfo{journal}{J.~Phys.~B} \textbf{\bibinfo{volume}{39}},
  \bibinfo{pages}{R127} (\bibinfo{year}{2006}).

\bibitem[{\citenamefont{Mayol et~al.}(2005)\citenamefont{Mayol, Ancilotto,
  Barranco, B{\"u}nermann, Pi, and Stienkemeier}}]{Mayol:2005}
\bibinfo{author}{\bibfnamefont{R.}~\bibnamefont{Mayol}},
  \bibinfo{author}{\bibfnamefont{F.}~\bibnamefont{Ancilotto}},
  \bibinfo{author}{\bibfnamefont{M.}~\bibnamefont{Barranco}},
  \bibinfo{author}{\bibfnamefont{O.}~\bibnamefont{B{\"u}nermann}},
  \bibinfo{author}{\bibfnamefont{M.}~\bibnamefont{Pi}}, \bibnamefont{and}
  \bibinfo{author}{\bibfnamefont{F.}~\bibnamefont{Stienkemeier}},
  \bibinfo{journal}{J. Low Temp. Phys.} \textbf{\bibinfo{volume}{138}},
  \bibinfo{pages}{229} (\bibinfo{year}{2005}).

\bibitem[{\citenamefont{Dalfovo}(1994)}]{Dalfovo:1994}
\bibinfo{author}{\bibfnamefont{F.}~\bibnamefont{Dalfovo}}, \bibinfo{journal}{Z.
  Phys. D} \textbf{\bibinfo{volume}{29}}, \bibinfo{pages}{61}
  (\bibinfo{year}{1994}).

\bibitem[{\citenamefont{Ancilotto et~al.}(1995)\citenamefont{Ancilotto,
  DeToffol, and Toigo}}]{Ancilotto:1995}
\bibinfo{author}{\bibfnamefont{F.}~\bibnamefont{Ancilotto}},
  \bibinfo{author}{\bibfnamefont{G.}~\bibnamefont{DeToffol}}, \bibnamefont{and}
  \bibinfo{author}{\bibfnamefont{F.}~\bibnamefont{Toigo}},
  \bibinfo{journal}{Phys. Rev. B} \textbf{\bibinfo{volume}{52}},
  \bibinfo{pages}{16125} (\bibinfo{year}{1995}).

\bibitem[{\citenamefont{Stienkemeier et~al.}(1995)\citenamefont{Stienkemeier,
  Higgins, Ernst, and Scoles}}]{Stienkemeier2:1995}
\bibinfo{author}{\bibfnamefont{F.}~\bibnamefont{Stienkemeier}},
  \bibinfo{author}{\bibfnamefont{J.}~\bibnamefont{Higgins}},
  \bibinfo{author}{\bibfnamefont{W.~E.} \bibnamefont{Ernst}}, \bibnamefont{and}
  \bibinfo{author}{\bibfnamefont{G.}~\bibnamefont{Scoles}},
  \bibinfo{journal}{Phys. Rev. Lett.} \textbf{\bibinfo{volume}{74}},
  \bibinfo{pages}{3592} (\bibinfo{year}{1995}).

\bibitem[{\citenamefont{Stienkemeier et~al.}(1996)\citenamefont{Stienkemeier,
  Higgins, Callegari, Kanorsky, Ernst, and Scoles}}]{Stienkemeier:1996}
\bibinfo{author}{\bibfnamefont{F.}~\bibnamefont{Stienkemeier}},
  \bibinfo{author}{\bibfnamefont{J.}~\bibnamefont{Higgins}},
  \bibinfo{author}{\bibfnamefont{C.}~\bibnamefont{Callegari}},
  \bibinfo{author}{\bibfnamefont{S.~I.} \bibnamefont{Kanorsky}},
  \bibinfo{author}{\bibfnamefont{W.~E.} \bibnamefont{Ernst}}, \bibnamefont{and}
  \bibinfo{author}{\bibfnamefont{G.}~\bibnamefont{Scoles}},
  \bibinfo{journal}{Z. Phys. D} \textbf{\bibinfo{volume}{38}},
  \bibinfo{pages}{253} (\bibinfo{year}{1996}).

\bibitem[{\citenamefont{Callegari et~al.}(1998)\citenamefont{Callegari,
  Higgins, Stienkemeier, and Scoles}}]{Callegari:1998}
\bibinfo{author}{\bibfnamefont{C.}~\bibnamefont{Callegari}},
  \bibinfo{author}{\bibfnamefont{J.}~\bibnamefont{Higgins}},
  \bibinfo{author}{\bibfnamefont{F.}~\bibnamefont{Stienkemeier}},
  \bibnamefont{and} \bibinfo{author}{\bibfnamefont{G.}~\bibnamefont{Scoles}},
  \bibinfo{journal}{J. Phys. Chem. A} \textbf{\bibinfo{volume}{102}},
  \bibinfo{pages}{95} (\bibinfo{year}{1998}).

\bibitem[{\citenamefont{B{\"u}nermann et~al.}(2007)\citenamefont{B{\"u}nermann,
  Droppelmann, Hernando, Mayol, and Stienkemeier}}]{Buenermann:2007}
\bibinfo{author}{\bibfnamefont{O.}~\bibnamefont{B{\"u}nermann}},
  \bibinfo{author}{\bibfnamefont{G.}~\bibnamefont{Droppelmann}},
  \bibinfo{author}{\bibfnamefont{A.}~\bibnamefont{Hernando}},
  \bibinfo{author}{\bibfnamefont{R.}~\bibnamefont{Mayol}}, \bibnamefont{and}
  \bibinfo{author}{\bibfnamefont{F.}~\bibnamefont{Stienkemeier}},
  \bibinfo{journal}{J. Phys. Chem. A} \textbf{\bibinfo{volume}{111}},
  \bibinfo{pages}{12684} (\bibinfo{year}{2007}).

\bibitem[{\citenamefont{Bovino et~al.}(2008)\citenamefont{Bovino, Bodo,
  Yurtsever, and Gianturco}}]{Bovino:2008}
\bibinfo{author}{\bibfnamefont{S.}~\bibnamefont{Bovino}},
  \bibinfo{author}{\bibfnamefont{E.}~\bibnamefont{Bodo}},
  \bibinfo{author}{\bibfnamefont{E.}~\bibnamefont{Yurtsever}},
  \bibnamefont{and} \bibinfo{author}{\bibfnamefont{F.~A.}
  \bibnamefont{Gianturco}}, \bibinfo{journal}{J. Chem. Phys.}
  \textbf{\bibinfo{volume}{128}}, \bibinfo{pages}{224312}
  (\bibinfo{year}{2008}).

\bibitem[{\citenamefont{Aub{\"o}ck et~al.}(2007)\citenamefont{Aub{\"o}ck, Nagl,
  Callegari, and Ernst}}]{Auboeck:2007}
\bibinfo{author}{\bibfnamefont{G.}~\bibnamefont{Aub{\"o}ck}},
  \bibinfo{author}{\bibfnamefont{J.}~\bibnamefont{Nagl}},
  \bibinfo{author}{\bibfnamefont{C.}~\bibnamefont{Callegari}},
  \bibnamefont{and} \bibinfo{author}{\bibfnamefont{W.}~\bibnamefont{Ernst}},
  \bibinfo{journal}{J. Chem. Phys.} \textbf{\bibinfo{volume}{111}},
  \bibinfo{pages}{7404} (\bibinfo{year}{2007}).

\bibitem[{\citenamefont{Lang et~al.}(2008)\citenamefont{Lang, Winkler, Strauss,
  Grimm, and Denschlag}}]{Lang:2008}
\bibinfo{author}{\bibfnamefont{F.}~\bibnamefont{Lang}},
  \bibinfo{author}{\bibfnamefont{K.}~\bibnamefont{Winkler}},
  \bibinfo{author}{\bibfnamefont{C.}~\bibnamefont{Strauss}},
  \bibinfo{author}{\bibfnamefont{R.}~\bibnamefont{Grimm}}, \bibnamefont{and}
  \bibinfo{author}{\bibfnamefont{J.~H.} \bibnamefont{Denschlag}},
  \bibinfo{journal}{Phys. Rev. Lett.} \textbf{\bibinfo{volume}{101}},
  \bibinfo{pages}{133005} (\bibinfo{year}{2008}).

\bibitem[{\citenamefont{Vetchinkin et~al.}(1993)\citenamefont{Vetchinkin,
  Vetchinkin, Eryomin, and Umanskii}}]{Vetchinkin:1993}
\bibinfo{author}{\bibfnamefont{S.~I.} \bibnamefont{Vetchinkin}},
  \bibinfo{author}{\bibfnamefont{A.~S.} \bibnamefont{Vetchinkin}},
  \bibinfo{author}{\bibfnamefont{V.~V.} \bibnamefont{Eryomin}},
  \bibnamefont{and} \bibinfo{author}{\bibfnamefont{I.~M.}
  \bibnamefont{Umanskii}}, \bibinfo{journal}{Chem. Phys. Lett.}
  \textbf{\bibinfo{volume}{215}}, \bibinfo{pages}{11} (\bibinfo{year}{1993}).

\bibitem[{\citenamefont{Vetchinkin and Eryomin}(1994)}]{Vetchinkin:1994}
\bibinfo{author}{\bibfnamefont{S.~I.} \bibnamefont{Vetchinkin}}
  \bibnamefont{and} \bibinfo{author}{\bibfnamefont{V.~V.}
  \bibnamefont{Eryomin}}, \bibinfo{journal}{Chem. Phys. Lett.}
  \textbf{\bibinfo{volume}{222}}, \bibinfo{pages}{394} (\bibinfo{year}{1994}).

\bibitem[{\citenamefont{Averbukh and Perel'man}(1991)}]{Averbukh:1991}
\bibinfo{author}{\bibfnamefont{I.~S.} \bibnamefont{Averbukh}} \bibnamefont{and}
  \bibinfo{author}{\bibfnamefont{N.~F.} \bibnamefont{Perel'man}},
  \bibinfo{journal}{Sov. Phys. Usp.} \textbf{\bibinfo{volume}{34}},
  \bibinfo{pages}{572} (\bibinfo{year}{1991}).

\bibitem[{\citenamefont{Higgins et~al.}(1998)\citenamefont{Higgins, Callegari,
  Reho, Stienkemeier, Ernst, Gutowski, and Scoles}}]{Higgins:1998}
\bibinfo{author}{\bibfnamefont{J.}~\bibnamefont{Higgins}},
  \bibinfo{author}{\bibfnamefont{C.}~\bibnamefont{Callegari}},
  \bibinfo{author}{\bibfnamefont{J.}~\bibnamefont{Reho}},
  \bibinfo{author}{\bibfnamefont{F.}~\bibnamefont{Stienkemeier}},
  \bibinfo{author}{\bibfnamefont{W.~E.} \bibnamefont{Ernst}},
  \bibinfo{author}{\bibfnamefont{M.}~\bibnamefont{Gutowski}}, \bibnamefont{and}
  \bibinfo{author}{\bibfnamefont{G.}~\bibnamefont{Scoles}},
  \bibinfo{journal}{J. Phys. Chem. A} \textbf{\bibinfo{volume}{102}},
  \bibinfo{pages}{4952} (\bibinfo{year}{1998}).

\bibitem[{\citenamefont{Stienkemeier and Vilesov}(2001)}]{Stienkemeier:2001}
\bibinfo{author}{\bibfnamefont{F.}~\bibnamefont{Stienkemeier}}
  \bibnamefont{and} \bibinfo{author}{\bibfnamefont{A.~F.}
  \bibnamefont{Vilesov}}, \bibinfo{journal}{J. Chem. Phys.}
  \textbf{\bibinfo{volume}{115}}, \bibinfo{pages}{10119}
  (\bibinfo{year}{2001}).

\bibitem[{\citenamefont{Stienkemeier et~al.}(2000)\citenamefont{Stienkemeier,
  Wewer, Meier, and Lutz}}]{Stienkemeier:2000}
\bibinfo{author}{\bibfnamefont{F.}~\bibnamefont{Stienkemeier}},
  \bibinfo{author}{\bibfnamefont{M.}~\bibnamefont{Wewer}},
  \bibinfo{author}{\bibfnamefont{F.}~\bibnamefont{Meier}}, \bibnamefont{and}
  \bibinfo{author}{\bibfnamefont{H.~O.} \bibnamefont{Lutz}},
  \bibinfo{journal}{Rev. Sci. Instrum.} \textbf{\bibinfo{volume}{71}},
  \bibinfo{pages}{3480} (\bibinfo{year}{2000}).

\bibitem[{\citenamefont{Nagl et~al.}(2008{\natexlab{a}})\citenamefont{Nagl,
  Aub{\"o}ck, Hauser, Allard, Callegari, and Ernst}}]{Nagl_prl:2008}
\bibinfo{author}{\bibfnamefont{J.}~\bibnamefont{Nagl}},
  \bibinfo{author}{\bibfnamefont{G.}~\bibnamefont{Aub{\"o}ck}},
  \bibinfo{author}{\bibfnamefont{A.}~\bibnamefont{Hauser}},
  \bibinfo{author}{\bibfnamefont{O.}~\bibnamefont{Allard}},
  \bibinfo{author}{\bibfnamefont{C.}~\bibnamefont{Callegari}},
  \bibnamefont{and} \bibinfo{author}{\bibfnamefont{W.}~\bibnamefont{Ernst}},
  \bibinfo{journal}{Phys. Rev. Lett.} \textbf{\bibinfo{volume}{100}},
  \bibinfo{pages}{063001} (\bibinfo{year}{2008}{\natexlab{a}}).

\bibitem[{\citenamefont{Nagl et~al.}(2008{\natexlab{b}})\citenamefont{Nagl,
  Aub{\"o}ck, Hauser, Allard, Callegari, and Ernst}}]{Nagl_jcp:2008}
\bibinfo{author}{\bibfnamefont{J.}~\bibnamefont{Nagl}},
  \bibinfo{author}{\bibfnamefont{G.}~\bibnamefont{Aub{\"o}ck}},
  \bibinfo{author}{\bibfnamefont{A.}~\bibnamefont{Hauser}},
  \bibinfo{author}{\bibfnamefont{O.}~\bibnamefont{Allard}},
  \bibinfo{author}{\bibfnamefont{C.}~\bibnamefont{Callegari}},
  \bibnamefont{and} \bibinfo{author}{\bibfnamefont{W.}~\bibnamefont{Ernst}},
  \bibinfo{journal}{J. Chem. Phys.} \textbf{\bibinfo{volume}{128}},
  \bibinfo{pages}{154320} (\bibinfo{year}{2008}{\natexlab{b}}).

\bibitem[{\citenamefont{Spiegelmann et~al.}(1989)\citenamefont{Spiegelmann,
  Pavolini, and Daudey}}]{Spiegelmann:1989}
\bibinfo{author}{\bibfnamefont{F.}~\bibnamefont{Spiegelmann}},
  \bibinfo{author}{\bibfnamefont{D.}~\bibnamefont{Pavolini}}, \bibnamefont{and}
  \bibinfo{author}{\bibfnamefont{J.-P.} \bibnamefont{Daudey}},
  \bibinfo{journal}{J. Phys. B} \textbf{\bibinfo{volume}{22}},
  \bibinfo{pages}{2465} (\bibinfo{year}{1989}).

\bibitem[{\citenamefont{Jeung}(1997)}]{Jeung:1997}
\bibinfo{author}{\bibfnamefont{G.-H.} \bibnamefont{Jeung}},
  \bibinfo{journal}{J. Mol. Spec.} \textbf{\bibinfo{volume}{182}},
  \bibinfo{pages}{113} (\bibinfo{year}{1997}).

\bibitem[{\citenamefont{LeRoy}(1995)}]{level}
\bibinfo{author}{\bibfnamefont{R.}~\bibnamefont{LeRoy}},
  \emph{\bibinfo{title}{Chemical physics research report}},
  \bibinfo{howpublished}{University of Waterloo, CP-555, 1995}
  (\bibinfo{year}{1995}).

\bibitem[{\citenamefont{Beuc et~al.}(2007)\citenamefont{Beuc, Movre, Horvatic,
  Vadla, Dulieu, and Aymar}}]{Beuc:2007}
\bibinfo{author}{\bibfnamefont{R.}~\bibnamefont{Beuc}},
  \bibinfo{author}{\bibfnamefont{M.}~\bibnamefont{Movre}},
  \bibinfo{author}{\bibfnamefont{V.}~\bibnamefont{Horvatic}},
  \bibinfo{author}{\bibfnamefont{C.}~\bibnamefont{Vadla}},
  \bibinfo{author}{\bibfnamefont{O.}~\bibnamefont{Dulieu}}, \bibnamefont{and}
  \bibinfo{author}{\bibfnamefont{M.}~\bibnamefont{Aymar}},
  \bibinfo{journal}{Phys. Rev. A} \textbf{\bibinfo{volume}{75}},
  \bibinfo{pages}{032512} (\bibinfo{year}{2007}).

\end{thebibliography}

\end{document}